\documentclass[showpacs,aps,prb,reprint,superscriptaddress ]{revtex4-1}
\usepackage{bm}
\usepackage{graphicx}
\usepackage{graphics}
\usepackage{amsmath,amssymb,amstext}
\usepackage{amsfonts}

\usepackage{epstopdf}
\usepackage{hyperref}
\hypersetup{
 colorlinks,%
 citecolor=blue,%
 linkcolor=blue,%
 urlcolor=blue
}
\usepackage{tikz}\usetikzlibrary{petri}

\begin{document}

\newcommand{\up}{ \mathbb{\uparrow}}
\newcommand{\down}{ \mathbb{\downarrow}}
\newcommand{\Gmat}{ \mathbb{G}}
\newcommand{\deltaG}{ \Delta \mathbb{G}}
\newcommand{\bsigma}{\boldsymbol{\sigma}}
\newcommand{\Beff}{\textbf{B}_{\textbf{eff}}}

\newcommand{\ket}[1]{\vert#1\rangle }
\newcommand{\bra}[1]{\langle#1\vert}

%%%%%%%%%%%%%%%%%%%%%%%%%%%%%%%%%%%%%%%%%%%%%%%%%%%%%%%%%%%%%%%%%%%%%%%%%%%%%%%
\title{Path integral simulation of exchange interactions in CMOS spin qubits }

\author{Jesus D. Cifuentes} 
\email{j.cifuentes_pardo@unsw.edu.au}
\affiliation{School of Electrical Engineering and Telecommunications, University of New South Wales, Sydney 2052, Australia}

%Geometry generation and modeling
\author {Philip Y. Mai} %p.mai@unsw.edu.au
\affiliation {School of Electrical Engineering and Telecommunications, University of New South Wales, Sydney 2052, Australia}

\author {Fr\'ed\'eric Schlattner} %
\affiliation {School of Electrical Engineering and Telecommunications, University of New South Wales, Sydney 2052, Australia}
\affiliation {Solid State Physics Laboratory, Department of Physics, ETH Zurich, 8093 Zurich, Switzerland.}

\author{H.\ Ekmel Ercan}
%\email{ekmelercan@gmail.com}
\affiliation{Department of Electrical and Computer Engineering, University of California, Los Angeles, Los Angeles, CA 90095, USA}

\author{MengKe Feng}%\email{mengke.feng@unsw.edu.au}
\affiliation {School of Electrical Engineering and Telecommunications, University of New South Wales, Sydney 2052, Australia}

\author{Christopher C. Escott} %\email{c.escott@unsw.edu.au}
\affiliation{School of Electrical Engineering and Telecommunications, University of New South Wales, Sydney 2052, Australia}
\affiliation{Diraq, University of New South Wales, Sydney 2052, Australia}

\author{Andrew S. Dzurak} 
\affiliation{School of Electrical Engineering and Telecommunications, University of New South Wales, Sydney 2052, Australia}
\affiliation{Diraq, University of New South Wales, Sydney 2052, Australia}

\author {Andre Saraiva}
\email{a.saraiva@unsw.edu.au}
\affiliation {School of Electrical Engineering and Telecommunications, University of New South Wales, Sydney 2052, Australia}
\affiliation{Diraq, University of New South Wales, Sydney 2052, Australia}

\date{ \today }

\begin{abstract}

The boom of semiconductor quantum computing platforms created a demand for computer-aided design and fabrication of quantum devices. Path integral Monte Carlo (PIMC) can have an important role in this effort because it intrinsically integrates strong quantum correlations that often appear in these multi-electron systems. In this paper we present a PIMC algorithm that estimates exchange interactions of three-dimensional electrically defined quantum dots. We apply this model to silicon complementary metal-oxide-semiconductor (MOS) devices and we benchmark our method against well-tested full configuration interaction (FCI) simulations. As an application, we study the impact of a single charge trap on two exchanging dots, opening the possibility of using this code to test the tolerance to disorder of CMOS devices. This algorithm provides an accurate description of this system, setting up an initial step to integrate PIMC algorithms into development of semiconductor quantum computers.
\end{abstract} 
%\pacs{ APS does not use it anymore}
%\keywords{Spin qubits, exchange coupling, path integral, Monte Carlo}

\maketitle
%%%%%%%%%%%%%%%%%%%%%%%%%%%%%%%%%%%%%%%%%%%%%%%%%%%%%%%%%%%%%%%%%%%%%%%%%%%%%%%

\section{\label{sec:Intro} Introduction}

Silicon spin qubits are rapidly emerging as one of the top contenders for quantum computing. Their similarities with CMOS transistors are fueling expectations of having a fully integrated quantum processor with millions of qubits, as required by current fault-tolerance thresholds\cite{beverland_assessing_2022,gidney_how_2021}.

 \begin{figure}[t]
 \centering
 \includegraphics[width=\linewidth]{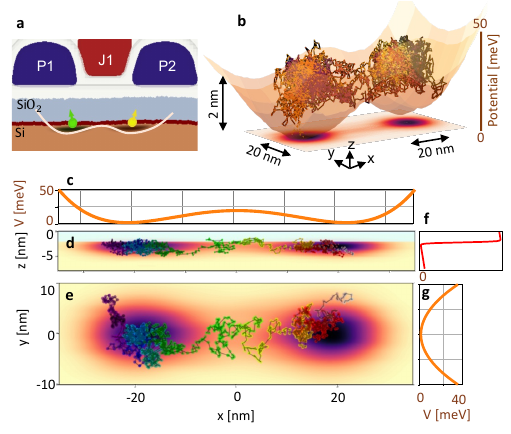}
 \caption{ \textbf{a,} Schematic of CMOS double quantum dot device. The quantum dots are formed around the two potential minima, below the oxide layer. \textbf{b,} Path integral simulation. The orange profile depicts the potential in the $x-y$ plane at $z = 0$. The x-y plane shows the electron path density. \textbf{c-g}, Comparison between electron density of the double quantum dot, with a single PIMC sampling and with the potential profile at each axis. \textbf{c}, Cut along the x-axis of the electrostatic potential. \textbf{d-e}, Electron density in the $xz$ (\textbf{d}) and in the $xy$ (\textbf{e}) planes. The change in color in the electron paths indicates the shift in the imaginary time. \textbf{f,g}, Potential profile along the $z$-axis (\textbf{f}) and the $y$-axis (\textbf{g}). The potential in the x-axis is at different scales. The large step of $3100$mV represents the gap in the conduction band between Si and SiO2.}
 \label{fig1}
\end{figure}

With the technology still at its dawn, it is necessary to guarantee that the key quantum operations will be repeatable and efficient across devices. One of these key operations is the exchange, which spin qubits rely on to execute entangling gates\cite{loss_quantum_1998,Li2010}. This interaction is activated when two spins are close enough to cause their wavefunctions to overlap. During the execution of a quantum algorithm, qubits should be continuously adjusted from an exchange OFF mode for single-qubit gates to an exchange ON mode for two-qubit gate operation. 

 Since the first proposal of this model in 1998~\cite{loss_quantum_1998}, a variety of quantum dot spin qubit technologies has emerged in semiconducting systems like silicon and germanium~\cite{burkard_semiconductor_2021}. Despite this, achieving repeatable and controllable exchange coupling is a difficult problem that all of these platforms have tackled with different levels of success. In the most successful ones, the implementation of two qubit gates followed soon after the observation of exchange interactions~\cite{hendrickx_four-qubit_2021, hendrickx_fast_2020,zajac_resonantly_2018}, with confirmed realizations of high fidelity two qubit gates ($>$99\%) in spin qubits in silicon~\cite{xue_quantum_2022,mills_two-qubit_2022, huang_fidelity_2019, tanttu2023stability}.

%exchange and architecture
The exchange coupling depends exponentially on the separation between quantum dots~\cite{Saraiva2007, Li2010}. That means that if the wavefunctions are too small or too distant from each other, or if they are affected by destructive Bloch oscillations in the lattice~\cite{Koiller2002}, the total overlap might be too small for exchange to be observed. This is probably the main reason for the success of gate-based quantum dots in this matter. Gate-defined dots are relatively large (10 to 100nm) and their size and position can be controlled electrically. Even more, in the last few years, interestitial exchange control gates between neighbouring dots have been implemented in quantum dot devices with the objective of accurately controlling the interdot barrier~\cite{Saraiva2022} (see FIG.~\ref{fig1}.\textbf{a}). This adaptation has significantly improved the success of these devices in creating controllable quantum entanglement across multiple platforms~\cite{geyer_silicon_2021, petit_universal_2020, petit_high-fidelity_2020,mills_two-qubit_2022, takeda_quantum_2022,Leon2021}. Now, with more and more devices having large and controllable exchange interactions, the pursuit is for optimization, extensive repeatability, and tolerance to disorder\cite{shehata2022modelling,cifuentes2023bounds}. 

%Fabricating a device that can generate nice individual qubits, which that at the same time can be driven close enough to exchange is a strong engineering challenge that has pushed gate litography to the limits of miniaturization. 

%The most successful technologies in this are so far gate-based quantum dot technologies which have achieved controllable exchange coupling in multiple cases and are now pushing for optimization, extensive repeatability, and uniformity [citations]. 

%why PIMC
 With these objectives in mind, we developed an exchange estimation tool based on the path integral Monte Carlo~\cite{Ceperley1995, Yan_2017}(PIMC) approach, which is an ideal tool to aid in the fabrication of spin qubit devices~\cite{Niquet_2020}. The main advantage of this AB-initio approach is its ability to tackle strongly interacting systems. PIMC treats the electrons as point-like particles immersed in the 3D potential repelling each other by Coulomb interactions, meaning that there is no need to compute costly Coulomb integrals. In this setup, the code samples hundreds of random electron paths with close to minimum action employing a Metropolis algorithm. Quantum operators, such as the energy or the electron density, are estimated from the mean values among the simulated random paths. This makes the algorithm very suitable for extensive parallelization. Each PIMC simulation runs individually and with very little cost in memory and computing power. No communication is needed between processor cores, meaning that a large number of PIMC paths can be simulated in parallel in a computational cluster.

%\textbf{Hello Andre: here is the list of the things missing for this paper:
%}
%\begin{enumerate}
%  \item Literature review comparing PIMC with traditional interpretation of exchange
%  \item Comparison between Full CI and PIMC
%  \item Send an email to the Full CI people. 
%  \item I have already included the PIMC Framework an a section about optimizing qubit architectures that replaces the trap calculations. I had to make modifications in the introduction, abstract and conclusions. They are in bold. 
%\end{enumerate}

%THis pape
In this paper, we use this approach to perform exchange coupling simulations in realistic 3D models of silicon CMOS double quantum dots. These dots are confined electrically against the Si/SiO$_2$ interface by the upper metallic gates observed in FIG.~\ref{fig1}.\textbf{a}. The exchange is controlled with the J-gate in the middle of two plunger gates (P1 and P2). To simulate this system, our PIMC code samples 500 realizations of two-electron paths inside the double quantum dot shown in FIG.~\ref{fig1}.\textbf{b}. Then building on top of the original approaches by Ceperley\cite{Ceperley1995} and Pedersen~\cite{Pedersen2010}, we sample paths that can exchange several times between the dots which allows us to estimate the exchange interaction from the relative increase in the total energy.

We observed the expected exponential dependence of exchange \textit{versus} interdot distance\cite{Saraiva2007, Li2010}, and compared it with a well-established full configuration interaction (FCI) approach. Then, we proceeded to demonstrate one of the main applications of this software, which is understanding the potential impact of impurities on this operation. Here, we show how a single negatively charged interface trap
can impact the two-dot system in different ways depending on the position where it is placed. 

This approach is extendable to other sources of disorder that are typical in CMOS technology. We have already used it, for instance, to understand the impact of Si/SiO$_2$ roughness on the exchange coupling, where we tested this method against actual experimental data \cite{cifuentes2023bounds}. A deep understanding of these sources of variability,  is essential in the design of realistic strategies to tolerate disorder and scale semiconductor quantum technologies\cite{Martinez_Variability_2022}.

%Then, building on top of original approaches, we induce the electrons to follow paths that exchange between the quantum dots and observed a linear increase in the action \textit{versus} the number of crossings. This allows us to estimate the exchange coupling

\section{Model of a CMOS double quantum dot (DQD)}
In general, the exchange coupling in semiconductors can be affected by Bloch oscillations in the lattice. This could be important in materials like silicon, in which there is a 6-fold valley degeneracy. However, in CMOS qubits the asymmetric confinement of the quantum dot against the (001)-interface lifts four of these degeneracies leaving only the two valley states in the z-axis\cite{Zwanenburg2013,Saraiva2009}. This is very convenient for CMOS, as the remaining Bloch oscillations are perpendicular to the in-plane orientation at which the exchange is controlled. While valley interference might still be a hurdle in CMOS quantum dots~\cite{tariq_impact_2022}, its impact is much smaller than in other technologies like donor qubits\cite{wang_highly_2016,Calderon_donor_2007,Koiller2002} and can be compensated with J-gate tunings. Because of this, in this initial approach we ignore the valley physics and focus on the effects of the architecture and J-gate tunability. 

In this work, we employ an effective mass approximation in which the full interacting Hamiltonian for a 2-electron double quantum dot is given by
\begin{equation}
\label{eq:Ht}
\begin{aligned}
 H(r_1(t), r_2(t)) = & \frac{1}{2} \vec{v}_{1}^\dagger M_{\text{Si}} \vec{v}_{1} + \frac{1}{2} \vec{v}_{2}^\dagger M_{\text{Si}}\vec{v}_{2} 
 \\ & + \frac{e^{2}}{4\pi\epsilon_{\rm Si}\left|\vec{r}_{1}-\vec{r}_{2}\right|} 
 \\ & + V_{\mathrm{DQD}}\left(\vec{r}_{1}\right)+V_{\mathrm{DQD}}\left(\vec{r}_{2}\right)
\end{aligned}
\end{equation}
where $M_{\text{Si}} = diag(0.19, 0.19, 0.98)m_e$ is a diagonal matrix with the effective mass of a silicon electron at each lattice orientation, and $\epsilon_{m}$ is the electrical permittivity of the material. Here we use the permittivity of silicon which is $\epsilon_{\text{Si}} = 11.7 \epsilon_0$. 
%However other sources use an average between the permitivity of Si and SiO$_2$ ($\epsilon_{\text{m}}=7.5$). %, for which the exchange would be slightly smaller. 
The potential of the 3D double quantum dot well is described by a model potential $V_{\mathrm{DQD}}$. The most accurate way to estimate this term is by performing electrostatic simulations of realistic qubit architectures with the tools available in COMSOL. For this first part of the paper, we use a simple quartic potential model in the $x$ axis two form the double quantum well (see FIG.~\ref{fig1}.\textbf{c}):

\begin{equation}
 \label{eq:VDQD}
 \begin{aligned}
 V_{DQD}(x,y,z) = & c_xx_L^2 x_R^2 - b_xd_{J}\left( x_L^2 + x_R^2 \right) \\ 
 & + \omega_y y^2 - zE_z + V_{\text{step}}\sigma(z),
 \end{aligned}
\end{equation}
\noindent where 
\begin{equation}
 x_L = x -\frac{d_{J}}{2} \ , \ x_R = x +\frac{d_{J}}{2} 
\end{equation}
and $d_{J}$ [nm] is a physical variable of the model that we associate with a relative interdot distance. In addition, in the other directions, the electrons are confined by a parabolic potential in the $y$ axis and an electric field $E_z$ in the $z$ axis (see FIG.~\ref{fig1}.\textbf{c-g}). We represent this barrier in FIG.~\ref{fig1}.\textbf{f}) as a soft step with height $V_{\text{step}}= 3.1$~eV mimicking the free conduction band offset between Si and SiO$_2$ multiplied by a sigmoid function
\begin{equation}
 \sigma(z)= \frac{1}{(e^{-4(z+2)/a_0} + 1)}
 \label{eq:sigma}
\end{equation}
at $z = -2$~nm, where $a_0 = 0.543$~nm is the silicon lattice parameter. 

\begin{figure*}%[ht!]
 \centering
 \includegraphics[width=\textwidth]{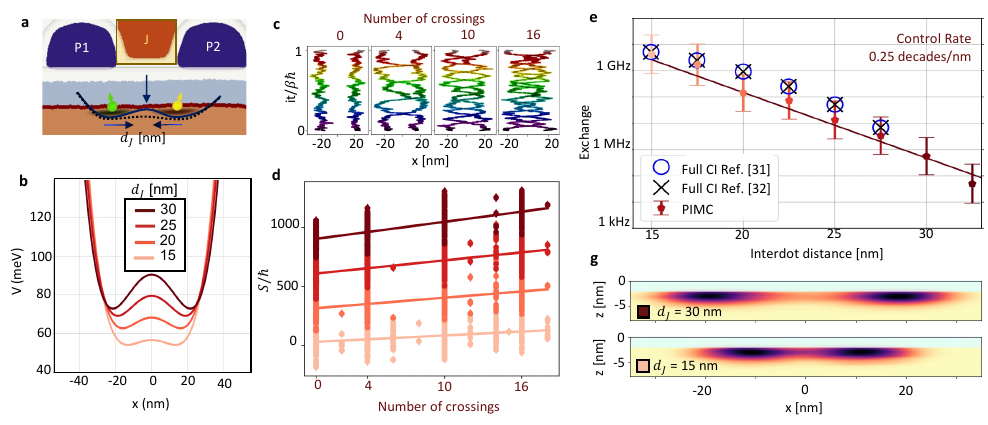}
 \caption{ \textbf{a}, Scheme of the operation of an exchange J gate. When the J gate is tuned the inter-dot barrier falls bringing the dots close enough to create exchange interactions. \textbf{b}, Cross section of the x-axis of the potential in equation~\eqref{eq:VDQD} for four values of $d_{J}$. \textbf{c}. Sampling of two-electron paths with $0$, $4$, $10$ and $16$ crossings. The color in the paths represents the variation in the imaginary time $it$, for comparison with Fig.~\ref{fig1}.\textbf{d-e}. \textbf{d}, Action $S$~\eqref{eq:action} \textit{versus} a number of crossings for the potentials in \textbf{b}. \textbf{e}, Exponential dependence of the exchange coupling \textit{versus} interdot distance. We benchmark PIMC results with full CI codes\cite{anderson_high-precision_2022, ercan_strong_2021}. \textbf{f}, Position density of the electron paths with $4$ crossings for $d_{J}= 15$~nm and $d_{J} = 30$~nm. }
 \label{fig2}
\end{figure*}

For a better approximation to realistic CMOS devices, we fitted this model to 
 potentials simulated in COMSOL for state-of-the-art devices obtaining the following approximate values for these parameters: $E_z= 20$~meV/nm, $\omega_y = 0.3$~meV/nm$^2$, $c_x = 8.1 \times 10^{-5}$~meV/nm$^4$ and $b_x = 6.48 \times 10^{-4} $~meV/nm$^3$. The only variable that we are going to sweep is $d_{J}$, which is designed to emulate the impact of a J gate. When J is pulsed on, the interdot distance $d_J$ becomes smaller which at the same time increases the exchange interaction (see FIG.~\ref{fig2}.\textbf{a}). Moreover, because $d_J$ is also multiplied by the $b_x$ term in equation~\eqref{eq:VDQD}, the interdot potential barrier decreases when J is pulsed (see FIG.~\ref{fig2}.\textbf{b}). This is confirmed in COMSOL simulations \cite{cifuentes2023bounds}. 

\section{\label{sec:Intro} Path Integral Monte Carlo (PIMC)}

%Path Integral Monte Carlo (PIMC) is a computational expresses the propagator of a quantum mechanical system as a sum over all possible paths between the initial and final states of the system.
PIMC has multiple applications across physics and chemistry \cite{Yan_2017}. As such, there is extended literature about this theory including instructions \cite{Westbroek_2018}, methods \cite{Ceperley1995}, and limitations \cite{Dornheim2019}. It has also been applied with notable success to the simulation of ideal multi-electron quantum dot systems \cite{Weiss2005,Kylanpaa2017}, including estimates of inter-dot exchange coupling in 2D dots \cite{Pedersen2010}. However, it does not yet exist, to our knowledge, a work that incorporates the complexity of realistic 3D quantum dot devices with the capacity of providing feedback to the process fabrication of semiconductor quantum architectures. This is the gap we are trying to fill. Here we summarize some of the most important concepts for this paper and define the notation that we are going to use.

Lets consider a time-independent Hamiltonian $\hat{H}$ with kinetic ($\hat{K}$), potential ($\hat{V}$) and interacting ($\hat{I}$) parts
\begin{equation}
    \hat{H} = \hat{K}+\hat{V}+ \hat{I}, 
\end{equation}
such as the one in equation~\eqref{eq:Ht}. The quantum evolution of a particle $\vert  \vec{r}, t\rangle$ is described by the Schrodinger equation 
\begin{equation}
i \hbar \frac{\partial}{\partial t} \vert \vec{r}, t\rangle=\hat{H} \vert  \vec{r}, t\rangle    \end{equation}
\noindent solved as the unitary evolution
\begin{equation}
    \psi(\vec{r},t)=e^{\frac{-i}{\hbar}\hat{H} t } \psi(\vec{r},0).
\end{equation}

The Path Integral formulation divides this unitary operator in infinitesimal time slices via Trotter's decomposition
\begin{equation}
\begin{aligned}
\label{eq:trot2}
       e^{\frac{-i}{\hbar}\hat{H} t} &= \lim\limits _{N\to\infty}\left(e^{\frac{-i}{\hbar}\hat{H}\tau} \right)^{N} \\
       &= \lim\limits_{N \to \infty} \left( e^{\frac{-i}{\hbar} \hat{K}\tau  } e^{\frac{-i}{\hbar} \frac{t}{N} \hat{V} }  e^{\frac{-i}{\hbar} \frac{t}{N} \hat{I} } \right)^N.     
\end{aligned}
\end{equation}

\noindent where $\tau := \frac{t}{N}$. The last step relies on the approximation ${e^{\tau (A+B)}=e^{\tau A}e^{\tau B} }$ which is exact when $\tau$ is small. After this, we can estimate the propagator of a particle between positions $\vec{r}_{0}$ and $\vec{r}_{N}$ as all possible sequences of these infinitesimal propagators that take the particle from the initial to the end point
\begin{equation}
\label{eq:trotterization}
\begin{aligned}
    \langle\vec{r}_{N},t\vert\vec{r}_{0},0\rangle=\langle\vec{r}_{N}\vert e^{-i\hat{H}t}\vert\vec{r}_{0}\rangle
    = \sum_{{\vec{r}_{j}}\in\mathbb{R}^{3\times N}}\prod_{j=0}^{N-1}\langle\vec{r}_{j+1}\vert e^{-iH\tau}\vert\vec{r}_{j}\rangle.
\end{aligned}
\end{equation}

For $N$ sufficiently large, the operators $ e^{\frac{-i}{\hbar} \frac{t}{N} \hat{K}  } $, $ e^{\frac{-i}{\hbar} \frac{t}{N} \hat{V} } $ and $ e^{\frac{-i}{\hbar} \frac{t}{N} \hat{I}}$ in~\eqref{eq:trot2} commute with each other (consequence of Baker–Campbell–Hausdorff formula), meaning that they can be applied directly to the wavefunctions in position space. This allows us to express~\eqref{eq:trotterization} as a compositon of the following propagators:

\begin{equation}
\begin{aligned}
    \langle\vec{r}_{j+1}\vert\hat{V}\vert\vec{r}_{j}\rangle & :=\frac{V(\vec{r}_{j})+V(\vec{r}_{j+1})}{2},
    \\ 
    \langle\vec{r}_{j+1}\vert\hat{K}\vert\vec{r}_{j}\rangle & :=\frac{m\vec{v}_{j}^{2}}{2}:=\frac{m\left\Vert \vec{r}_{j+1}-\vec{r}_{j}\right\Vert ^{2}}{2\tau^2}.
    \end{aligned}
\end{equation}
For 2 electron interactions, we would require a second index to describe the particle number.
\begin{equation}
    \langle\vec{r}_{1,j+1}\vert\hat{I}\vert\vec{r}_{2,j}\rangle  := \frac{1}{2}\frac{e^{2}}{4\pi\epsilon}\left( \frac{1}{\left|\vec{r}_{1,j}-\vec{r}_{2,j}\right|} + \frac{1}{\left|\vec{r}_{1,j+1}-\vec{r}_{2,j+1}\right|} \right).
\end{equation}

In total, the propagator can be estimated as 
\begin{equation}
\label{eq:trotter}
 \begin{aligned}
 \langle\vec{r}_{N},t\vert\vec{r}_{0},0\rangle
 & = \sum_{ \{ \vec{r} \}_j \in \mathbb{R}^3} e^{\frac{i}{\hbar} \mathcal{S}(\{ \vec{r} \}_j )},
 \end{aligned}
\end{equation}
where $\mathcal{S}(\{ \vec{r} \}_j)$ is the accumulated action over a path $\{ \vec{r} \}_j$ in the position space, such that 
\begin{equation}
 \label{eq:action}
 \mathcal{S}(\{ \vec{r} \}_j) = \sum_{j=0}^{N} \tau H\left( \vec{r}_j \right).
\end{equation}

One of the main aspects of this method is replacing $t$ by an imaginary time $i \beta/\hbar$.  When this is done, equation~\eqref{eq:trotter} gains an entire new significance as each individual term $ e^{\frac{-i}{\hbar} \mathcal{S}}$ is replaced by a Boltzman term $ e^{-\beta \sum_j H(\vec{r}_j)}$. This transformation creates a parallel between this unitary evolution and statistical mechanics where the variable $\beta$ can be thought as the inverse of a temperature $1/k_BT$. In this paper, we simulate electrons in temperatures down 1 K, which is equivalent to simulated total time lengths of 5 picoseconds. 

 %This means that temperature is indeed a variable in PIMC simulations and it is inversely proportioned to the parameter $t = i \beta/\hbar$. 

%Equation~\eqref{eq:trotter} becomes exact when $N_t$ is sufficiently large, and this discretization allows for the estimation of Z numerically $\mathcal{Z}$ numerically, with the method known a Path Integral Monte Carlo. 

In this new representation, we can think that the statistics of the operators are related to the electron paths $\{r\}_i$ which are distributed with a probability $e^{-\beta \sum_j H(\vec{r}_j)}$. Because of the exponential, only the electron paths that have a relatively small action are going to be relevant. PIMC makes an importance sampling of these paths employing a Metropolis algorithm. 

The metropolis sampling starts with a random trajectory which is to be optimized for minimal action $S$ through a series of random updates that are proposed after each iteration. At each one of these, the software proposes a modification to a section of the electron paths. Then, depending on its impact on the action, the software accepts or rejects the update according to the following rule. If the resulting action is smaller than before, it is always accepted. In contrast, if it is higher, the algorithm accepts the update with probability 
\begin{equation}
 p = e^{\frac{-\Delta S}{\hbar} },
\end{equation}
where $\Delta S$ is the difference between the new and the  old action. This last part is required to achieve a static balance in the algorithm~\cite{Ceperley1995}. Today, there exists a variety of updates used in PIMC algorithms (single slice, center of mass displacement, etc). For this paper, we chose a specific set of them which we described in the supplementary. 

Once the metropolis algorithm is implemented, it is possible to sample a varied set of random paths $\{ \vec{r} \}_i$ with relatively small action. The mean of an operator $\hat{\mathcal{O}}$ can be computed from the average of the output among the sampled random paths $\mathcal{P}$ \cite{Westbroek_2018}
\begin{equation}
 \langle \mathcal{O} \rangle = \frac{1}{N_R}\sum_{\{\vec{r}\}_i \in \mathcal{P}}  \langle \mathcal{O}(\vec{r}_i) \rangle 
\end{equation}
where $N_R$ is the total number of paths simulated. In addition, it is also possible to compute statistical errors $\Delta \mathcal{O}$ from the variance of operators as
\begin{equation}
\Delta \mathcal{O}= Z_{95\%} \frac{ \text{STD}(\mathcal{O})}{\sqrt{N_R}} = Z_{95\%} \sqrt{ \frac{ \langle \mathcal{O}^2 \rangle - \langle \mathcal{O} \rangle^2 }{N_R} } ,
\end{equation}
where $Z_{95\%}\approx 1.96$ is the z-score for the $95\%$ confidence interval. This allows us to estimate uncertainties in our computations. 

\section{Computation of the Exchange coupling with path integral Monte Carlo}

To simulate a system with two electrons we replaced in equation~\eqref{eq:action} the two electron Hamiltonian~\eqref{eq:Ht}. A visual representation of one of the sampled electron paths is observed in FIG.~\ref{fig1}.\textbf{b}. The bulk of the trajectories will be concentrated close to the minimums of the parabolic potentials, with certain paths crossing from one dot to the other. In addition to this, PIMC also provides a proper way to visualize the electron density. This can be done by creating a histogram of the position of the electrons over all realizations. The result is shown in FIG.~\ref{fig1}.\textbf{d-e} and compared with the potential profile in the different axes.

To compute the exchange coupling explicitly, we build on top of the original approach of Pedersen \textit{et al.}~\cite{Pedersen2010} in two dimensions. Their method is based on a type of bosonization of the paths. Traditionally, the simulation of fermionic paths requires a consideration of all possible path-exchanging electrons, which gain a negative sign in their action upon exchange and lead to what is known as the sign problem. In the special case of only two electrons, however, one is able to break down the time evolution (or, equivalently, the partition function) into paths that result in an even or odd number of exchanges (considering spins as completely separable from the orbital part of the wavefunction). Sampling the two types of paths separately as if they were bosonic particles and comparing them allows us to determine their energy difference. This reflects the difference in energy between singlets (spatially symmetric paths) and triplets (spatially anti-symmetric paths), which defines the two-particle exchange. This trick would fail in the most general case with either more electrons or if spin-orbit coupling made the breakdown between spin and orbital parts of the wavefunction impossible.

Then, the actual numerical calculation becomes very efficient by simulating two types of paths. The first type is when both electrons are confined below their own dot without exchanging. Let's call $S_0$ the average action for these paths. In the second type, the electrons are allowed to exchange a single time from one dot to the other and have an action that we call $S_1$. It is then expected that $S_1$ is larger than $S_0$ by an amount $\delta S$ because in $S_1$ the electrons are forced to pass through the interdot barrier that has a higher potential. This difference is related to the exchange coupling by
\begin{equation}
 e^{-\beta J} = \frac{ e^{-S_{0}/\hbar } - e^{-S_{1}/\hbar} }{e^{-S_{0}/\hbar } + e^{-S_1/\hbar}}.
\end{equation}
which means that 
\begin{equation}
 \label{eq:J}
 J = \frac{-1}{\beta} \ln \left( \frac{ e^{-S_{0}/\hbar } - e^{-S_1/\hbar} }{e^{-S_{0}/\hbar } + e^{-S_1/\hbar}} 
 \right)
 \approx \frac{2}{\beta} e^{-\delta S/\hbar},
\end{equation}
where the last approximation is valid as long as $e^{-\delta S/\hbar}$ is small, as we usually find in the simulations. These two states $S_0$ and $S_1$ can be associated with the spin singlet (symmetric wavefunction in position space) and spin-triplet state (anti-symmetric wavefunction in position space)~\cite{Ceperley1995} which correlates this method with the traditional interpretation of exchange coupling.

\begin{figure}[b]
 % \centering
 \includegraphics[width=\linewidth]{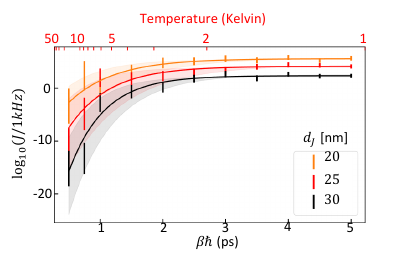}
 \caption{ Convergence of exchange estimate \textit{versus} time length $\beta \hbar$, which also sets represents the inverse of the temperature ${\beta=1/k_BT}$. }
 \label{fig3}
\end{figure}

\begin{figure*}[bt]
\centering
 \includegraphics[width=\linewidth]{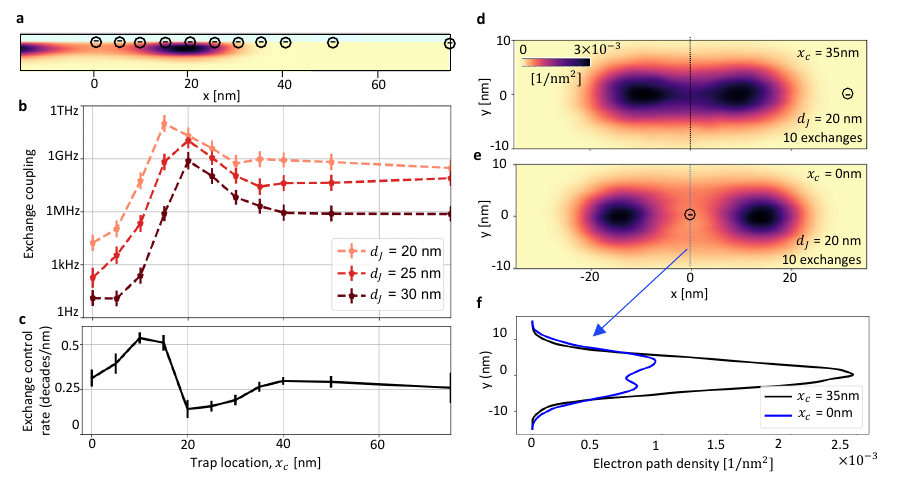}
 \caption{\textbf{a}, Schematic view of the position of the negative trap charges that are analyzed in these simulations. \textbf{b}, Exchange coupling \textit{vs} trap position. \textbf{c}, Impact of trap location on the rate of exchange control ($\frac{\text{d}\log_{10}(J)}{ \text{d} d_J}$) in decades per volt. \textbf{d-e}, Electron densities at $xc = 35$~nm and $xc = 0$~nm for PIMC paths that exchange 10 times between both quantum dots. $d_J =$~20 nm. \textbf{f}, Cut of the electron trap
density at the interdot channel for $x_c = 35$~nm and $x_c = 0$nm.}
 \label{fig4}
\end{figure*}

%In addition, because they electrons are ruled by fermion statistics the exchanging paths should have a negative contribution\cite{Ceperley1995}. 

While the initial results for 2D quantum dots were successful\cite{Pedersen2010}, implementing this idea for more realistic 3D silicon quantum dots turned out to be problematic as the statistical dispersion of the sampled paths measured, for instance, by their standard deviation $\sigma(S)$ was significantly higher than their difference $ \sigma(S_0) \approx \sigma(S_1) > \delta S$, making it hard to estimate $\delta S$ accurately. We solved this with a modification to the algorithm. Instead of just simulating paths that crossed one single time, we simulated paths that exchanged multiple times in the system. We verified that each exchange carried an additional constant value to the action, implying that $S_{N_c}$ increased by a linear rate with respect to the number of exchanges between the electrons $N_c$. This is observed in FIG.~\ref{fig2}.\textbf{d} in which we show the dispersion of $N_c$ \textit{versus} $S_{N_c}$ of $500$ paths simulated for each of the four potential configurations in FIG.~\ref{fig2}.\textbf{b}. The slope of each of these regressions gives and estimate for $\delta S$, from which we can compute the exchange coupling using equation~\eqref{eq:J}. This also provides a natural way to compute the error bars as the standard deviation of the slope in the linear regression multiplied, in this case, by 1.96 (the z-score associated to the 95\% confidence interval).

Figure~\ref{fig2}.\textbf{e} shows the output values of our exchange calculations. Notice that the exchange coupling decreases exponentially with the interdot distance as expected~\cite{Saraiva2007}. To ensure that our estimates were accurate enough we compared our results with two Full CI algorithms implemented independently \cite{anderson_high-precision_2022,ercan_strong_2021}. Details of Full CI calculations can be found in these references. 

%Further details about the full CI simulations can be found in the supplementary materials. 

%\textbf{We need to include them as coauthors right?.. should we send them another email? telling that the paper is coming. }.

We have a deeper look into what is happening in FIG.~\ref{fig2}.\textbf{g}. The plot compares the histogram of the position of the electrons for paths that exchanged four times in the system. While this metric is not the same as standard electron density in quantum mechanics, it is still useful to understand how the electrons distribute across the double dot when performing exchange. Notice that the density at the interdot region increases significantly when the $d_J$ decreases from $30$~nm to $15$~nm which contributes to a strong enhancement of the exchange coupling at a rate of 0.25 decades/nm. In particular, note that when the dots are more separated from each other, the exchange is as low as $10$~kHz. At this scale, the exchange is usually not visible in a standard qubit spectroscopy experiment as other effects such as disorder or spin-orbit coupling become dominant\cite{cifuentes2023bounds}. An important challenge for this architecture is to fabricate devices in which it is always possible to turn ON and OFF the exchange coupling consistently. And here we see that CMOS devices rely on this high tunability of the interdot distance to perform this operation. 

A final concern in this algorithm is the role of temperature which is inversely related to the variable $\beta$. In Fig.~\ref{fig3} we show that the exchange simulations converge for $\beta \hbar >$ 2~ps. At this point, the simulated temperature is lower than 2 Kelvin. As qubit measurements occur at temperatures ranging from tens of mK and up to $1$~K, we can assume that the temperature will not have a significant role in the value of the exchange coupling.

\section{Impact of static trapped charge on the system}

%Here we simulate an static trap charge \cite{Rahman2012}. Its effect will depend on its specfic location of the trap on the two dimensional plane, and how deep it is into the SiO$_2$. We can simulate this by including in the Hamiltonian of equation ~\eqref{eq:VDQD} the Coulomb repulsion from a trap
To test additional applications of this PIMC algorithm we make an initial approach to describe the impact of disorder on exchange interactions. Here we calculate the effect of a static charge trap by adding a Coulomb interaction term to our Hamiltonian, that describes the repulsion between the charge trap and the dot electrons as previously described in \cite{Rahman2012}. For each electron $i\in{1,2}$, we include in equation ~\eqref{eq:VDQD}
\begin{equation}
  H_{Trap}(\vec{r_i}) = \frac{1}{4\pi\epsilon_{Si}}\frac{e^2}{\vert \vec{r_i} - \vec{r}_c \vert }, 
\end{equation}
 \noindent where $\vec{r}_c = (x_c, y_c, z_c)$ is the position of the trap.

As the focus of this paper is only to show the potential of PIMC to tackle these problems, we limit this paper to the simulation of a single negative interface trap ($z_c$=~-1~nm, the same level as the SiO$_2$ oxide barrier) placed in the dot line ($y_c$=0~nm) that passes through the middle of both quantum dots. Here, $x_c$ is left as the only variable. This is already the worst-case scenario as any charge that is outside the dot-line or that is more deep into the oxide would have a smaller impact on the potential configuration.

We performed exchange simulations for traps located a the positions shown in FIG.\ref{fig4}~\textbf{a} and presented the results
in FIG.\ref{fig4}~\textbf{b} for diferent values of $d_J$ . Notice that when the electron is far enough ($x_c \approx 40$~nm) we recover the pristine simulation without any trap. In contrast, when the trap is slightly closer to the system we
can see that exchange increases or decreases depending on whether the trap is inside or outside the double quantum well. This occurs asymmetrically for the different values of $d_J$, which explains why there is also an impact on the exchange control rate $\frac{\text{d}\log_{10}(J)}{ \text{d} d_J}$ (see FIG.\ref{fig4}~\textbf{c}). All this makes sense because the negative trap pushes the dots closer together when it is outside of the double quantum well, while it drives them apart when it is inside (FIG.\ref{fig4}~\textbf{d-e}). The most critical scenario is when the trap is exactly inside the interdot channel. But even in this case, we can see that at $d_J = 20$~nm there is still an acceptable exchange coupling because of the existing electron density the interdot channel surrounding the negative trap (FIG.\ref{fig4}~.\textbf{e-f}).

\section{Prospects for path integral in the simulation of quantum dot qubits }

We have demonstrated that PIMC can be applied to the simulation of interacting effects on quantum dot qubits. However, our initial success with this protocol is in part because the electrons that we simulated lie in different quantum dots, and the paths only crossed each other when exchange is performed. That means that at the current stage we can perform multi-electron simulations as long as the electrons remain in separate dots for most of the time. 

Even with these constrains, this approach be used simulate quantum dots chains (or grids) which are of high interest in large-scale quantum computing. As long as the electrons do not lie in the same dot, PIMC is able to simulate all of them interacting with each other with only a linear impact on memory and complexity. This can be used to study inter-dot correlations, which could help to understand the crosstalk effects between electron charges at different dots.

For a more general perspective we would like to simulate systems in which multiple electrons can occupy the same quantum dot. This is very interesting for the field as it has been shown that it is possible to control spin qubits at the outer shell of multi-electron quantum dots, with possible improvements in the coherence of single qubits~\cite{Leon2020} and also on the strength of the exchange interactions between two qubits\cite{Leon2021}.

However, simulating multi-electron quantum dots can be problematic in PIMC due to the infamous fermion sign problem~\cite{Ceperley1995}. Despite this concern, it's noteworthy that methods to tackle this issue have significantly improved in recent years~\cite{Kylanpaa2017, Dornheim2019} with encouraging results in simulating 2D multi-electron quantum dots~\cite{Weiss2005}. Additionally, to fully simulate silicon dots, valley physics must be included in the model as in a well-closed shell structure, a third electron would occupy the upper valley state, and not the first p-orbital as usual~\cite{Leon2020,ercan_multielectron_2023}.

\section{Conclusions }

We demonstrated here a method to compute exchange coupling in realistic 3D silicon quantum dots, which can be applied to the optimization of device architectures and studies of tolerance of disorder in silicon qubits. Our results agreed with equivalent simulations with full configuration interaction algorithms, which are considered to be a current standard in simulating strongly correlated systems. We also showed that PIMC provides proper methods to visualize the electron density, thus allowing us to study features in the quantum dot structure. This is well observed in the trap simulation where the electron density curves around the negative trap.

We expect that this initial approach motivates the further applications of PIMC algorithms in semiconductor qubits, either by studying charge correlations in large grids of single electron quantum dots or by leveraging the code to simulate the exquisite physics of multi-electron spin qubits. If it is well combined with standard electrostatic simulation software such as COMSOL Multiphysics, PIMC algorithms could provide substantial support to the fabrication of optimal and highly repeatable CMOS spin qubit devices.

\section{Acknowledgements}
We thank Christopher Anderson, Mark Gyure, Mark Friesen, and Susan Coppersmith for useful discussions. We acknowledge support from the Australian Research Council (FL190100167 and CE170100012), the US Army Research Office (W911NF-23-1-0092), and the NSW Node of the Australian National Fabrication Facility. The views and conclusions contained in this document are those of the authors and should not be interpreted as representing the official policies, either expressed or
implied, of the Army Research Office or the US Government. The US Government is authorized to reproduce and distribute reprints for Government purposes notwithstanding any copyright notation herein. M.K.F., and J.D.C. acknowledge support from the Sydney Quantum Academy. This project was undertaken with the assistance of resources and services from the National Computational Infrastructure (NCI), which is supported by the Australian Government and includes computations using the computational cluster Katana supported by Research Technology Services at UNSW Sydney.

\section*{Appendix A: Path initialization: \label{sec:initialization}}

Each initial position of an electron in the $i^{th}$ quantum dot ($i \in \{1,2\}$) at time $t$ is initialized from a random sampling of the normal distribution $\mathcal(N)(x_i, \sigma_i) (t)$, where $x_i$ is the minimum of the potential of dot $i$ and $\sigma_i$ is chosen to be sufficiently large to cover for both dots. Here we chose $\sigma_i = 30$nm. We didn't observe a substantial dependence of this variable on the output of the algorithm as long as it is big enough to cover an important region around the dots. 

To simulate paths with multiple exchanges in the double quantum dot we alternate the position of the electrons during the imaginary time. For instance, to initialize an electron path with 4 exchanges, we can divide the time frame $\beta$ in four sections: i. $t < \beta/4 $, ii. $\beta/4<t<\beta/2$, iii. $\beta/2<t<3\beta/4$, iv. $3\beta/4<t<\beta$. In sections i. and iii. the first electron is sampled at the center of dot 1 and the second electron is sampled in dot 2. Instead, in sections ii. and iv. the first electron is initialized in dot 2 and the second electron is initialized in dot 1. This will guarantee that the electrons are most likely going to perform 4 exchanges after the simulation.

This is, however, not an strict rule. Some electron exchanges can disappear or emerge during the metropolis iteration of the PIMC simulation. To avoid that this happening so often that it becomes intractable the parity of the number of exchanges is protected during the PIMC simulations. This is done, by fixing periodic boundary conditions in the time axis ($\vert \vec{r}, t = 0\rangle = \vert \vec{r}, t = \beta \rangle $). By doing this, a path initialized to have 4 exchanges, for instance, can only end in a path with the same parity.

Because of this reason, changes in the number of electron exchanges during the PIMC simulation are not so common, and they are usually easy to track.  We implemented a quick algorithm during the post-processing to  read the sampled electron paths and estimate the real number of crossings after the simulation. As observed in FIG.~\ref{fig2}.\textbf{d} most of the paths coincide with one of the original number of crossings in the initialization (0, 4, 10, 16).  The remaining paths that do not coincide with this number, are those ones where the number of crossings changes during optimization of the PIMC paths. 

\section*{Appendix B: Updates} The current implementation only includes two types of updates in the simulation\cite{Ceperley1995} that provided the best configuration for our purposes:

\textbf{Staging update:} For a time step $t_i$ chosen randomly, the algorithm time slice subsection starting at $t_i$ and with a defined length of $T \geq 3$, such that it ends at $t_i + T$. The update replaces all middle positions $\textbf{r}_t'$ of the electron, with $t' \in (t_i +1, t_i + \delta T-1)$, by new positions sampled with a normal distribution $\mathcal{N}(\mu_{t'}, \sigma_t' )$ where

\begin{equation}
 \begin{aligned}
 \boldsymbol{r}_{t'} &=\frac{1}{T} \left[(t_i+T-t') r_{t}+(t'-t_i) \boldsymbol{r}_{t_i+T}\right]
 \\
 \sigma^{2} &=\frac{\tau \hbar}{2m} \frac{2}{\frac{1}{(t_i+T-t')
 \tau}+\frac{1}{(t'-t_i) \tau}}.
 \end{aligned}
\end{equation}
Here $m $ is the effective mass on the direction of motion. This update already covers for the convergence in the kinetic energy and then the acceptance criteria only checks for the difference in action attributed to the change in potential energy. Meaning that if the action increases the code accepts the update with probability
\begin{equation}
 p = e^{\frac{- \beta}{\hbar} (V(R_n(t_i,t_i +T)) -V(R_o(t_i,t_i +T)) )}
 \label{eq:acceptGauss}
\end{equation}
where $V(R_{o,n}(t_i,t_i +T))$ accounts for the potential energy between $t_i$ and $t_i+T$ of the old ($R_o$) and the new path($R_n$) path respectively. 
During the algorithm the length of the subpaths $T$ changes to obtain a better estimator for the kinetic energy. Thus, we initially set $T=27$ and when the algorithm reaches convergence $T$ is updated to $9$ and finally to $T=3$. This has a double purpose. At the beginning of the algorithm, it is necessary that the paths have a large range of movement to be able to explore varied types of paths. $T=27$ is ideal for this. When the algorithm converges, the estimate for the action will be more accurate if paths updates are finely tuned. This is done with $T=3$. The code switches between these modes.

\textbf{Center of mass update:} 
We also implement a center of mass update. It takes the entire path and moves it in the direction $r'$ where $r'$ is obtained from a random uniform distribution in the ranges $([-a_x,a_x ], [-a_y, a_y], [-a_z, a_z])$ where we set $a_x = a_y = 5nm$ and $a_z = 1nm$. The code is given a probability of $0.05$ of implementing this update, and the update is accepted according to the rule in equation~\eqref{eq:acceptGauss} as it does not involve a change in the kinetic energy. 

\begin{figure}%[ht!]
 % \centering
 \includegraphics[width=3.5in]{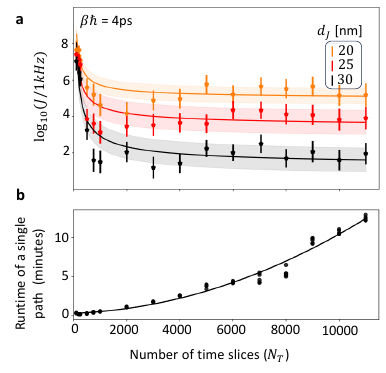}
 \caption{ Convergence of exchange $(J)$ simulations, for 500 PIMC sampling paths. \textbf{a} Convergence $J$ \textit{versus} number of time slices $N_t$ for $\beta \hbar = 4$ps. \textbf{b} Dependence of the runtime of individual path simulations \textit{versus} number of time slices. At about 8000 time slices where we run most of the simulations, the PIMC runtime of a single path is about 5 minutes. With 10 cores running in-parallel in a cluster, 500 paths can be simulated in 50 minutes.}
 \label{fig5}
\end{figure}

\section*{Appendix C: Convergence and optimization }

To obtain the results displayed in FIG.~\ref{fig2} we first had to verify for the convergence of the algorithm at low temperatures (high $\beta$) and number of time slices. We show in \ref{fig3}, that for for paths with 8000 time slices the PIMC exchange results get stable after $\beta \hbar > 2$ps which corresponds to temperatures lower than $7$K. In all cases, we computed exchange with 500 path samplings with the initialization equally distributed between 0, 4, 10 and 16 exchanges. In the first simulations $(\beta \hbar < 1ps)$ we observed that the time length was too small for the exchange number to be preserved. In consequence the final number of crossings of the output PIMC simulations was significantly lower than the initialized number. Hence most of the PIMC paths had either 0 or 2 crossings which contributed to a wrong estimate of the exchange coupling. This changes after $\beta \hbar = 2$ps when the time length is long enough for the electrons to exchange multiple times. 

Once we know that the algorithm converges for $\beta$, we also tested the Trotter convergence in $N_T$. Taking $\beta = 4$ps, we create FIG.~\ref{fig5}.\textbf{a} by simulating the convergence of the exchange coupling \textit{versus} the number of time slices. We can observe the exchange rate converges at around $5000$ time slices. As it commonly happens in other PIMC algorithms the error bars do not significantly increase with the number of time slices. This happens because the uncertainty in the exchange depends on the standard deviation of the slope of the linear regression of  $S$ \textit{versus} number of exchanges. This does not depend significantly on $N_T$. 

Also, simulating longer paths implies a longer runtime of the algorithm. This is shown in FIG.~\ref{fig5}.\textbf{b} which depicts the runtime of single PIMC simulations at different path discretizations. Then we performed a quadratic fit of the function showing that the runtime of the algorithm scales at $\sim N_t^2$. 

For this paper, we perform all of the simulations with 8000 time slices which accounts for a $5$minute runtime per path. The simulations were simulated with extensive parallelization in Katana(UNSW) and Gadi (NCI) clusters, each one with with low memory requirements $<1$MB and without any communication between multiple cores. This allowed us to perform large amounts of exchange simulations in an amount that is suitable for random variability studies (hundreds of simulations with varying parameters)\cite{cifuentes2023bounds}. 

There is also plenty of space for optimization in this code. It was fundamentally written in python, with proper vectorization, but could be improved systematically if written in C or C++. Optimizing the set of updates used during each path simulation and the number of paths sampled could also significantly improve the performance of the code.

\bibliography{PIMC}

\begin{thebibliography}{41}%
\makeatletter
\providecommand \@ifxundefined [1]{%
 \@ifx{#1\undefined}
}%
\providecommand \@ifnum [1]{%
 \ifnum #1\expandafter \@firstoftwo
 \else \expandafter \@secondoftwo
 \fi
}%
\providecommand \@ifx [1]{%
 \ifx #1\expandafter \@firstoftwo
 \else \expandafter \@secondoftwo
 \fi
}%
\providecommand \natexlab [1]{#1}%
\providecommand \enquote  [1]{``#1''}%
\providecommand \bibnamefont  [1]{#1}%
\providecommand \bibfnamefont [1]{#1}%
\providecommand \citenamefont [1]{#1}%
\providecommand \href@noop [0]{\@secondoftwo}%
\providecommand \href [0]{\begingroup \@sanitize@url \@href}%
\providecommand \@href[1]{\@@startlink{#1}\@@href}%
\providecommand \@@href[1]{\endgroup#1\@@endlink}%
\providecommand \@sanitize@url [0]{\catcode `\\12\catcode `\$12\catcode
  `\&12\catcode `\#12\catcode `\^12\catcode `\_12\catcode `\%12\relax}%
\providecommand \@@startlink[1]{}%
\providecommand \@@endlink[0]{}%
\providecommand \url  [0]{\begingroup\@sanitize@url \@url }%
\providecommand \@url [1]{\endgroup\@href {#1}{\urlprefix }}%
\providecommand \urlprefix  [0]{URL }%
\providecommand \Eprint [0]{\href }%
\providecommand \doibase [0]{http://dx.doi.org/}%
\providecommand \selectlanguage [0]{\@gobble}%
\providecommand \bibinfo  [0]{\@secondoftwo}%
\providecommand \bibfield  [0]{\@secondoftwo}%
\providecommand \translation [1]{[#1]}%
\providecommand \BibitemOpen [0]{}%
\providecommand \bibitemStop [0]{}%
\providecommand \bibitemNoStop [0]{.\EOS\space}%
\providecommand \EOS [0]{\spacefactor3000\relax}%
\providecommand \BibitemShut  [1]{\csname bibitem#1\endcsname}%
\let\auto@bib@innerbib\@empty
%</preamble>
\bibitem [{\citenamefont {Beverland}\ \emph {et~al.}(2022)\citenamefont
  {Beverland}, \citenamefont {Murali}, \citenamefont {Troyer}, \citenamefont
  {Svore}, \citenamefont {Hoefler}, \citenamefont {Kliuchnikov}, \citenamefont
  {Low}, \citenamefont {Soeken}, \citenamefont {Sundaram},\ and\ \citenamefont
  {Vaschillo}}]{beverland_assessing_2022}%
  \BibitemOpen
  \bibfield  {author} {\bibinfo {author} {\bibfnamefont {M.~E.}\ \bibnamefont
  {Beverland}}, \bibinfo {author} {\bibfnamefont {P.}~\bibnamefont {Murali}},
  \bibinfo {author} {\bibfnamefont {M.}~\bibnamefont {Troyer}}, \bibinfo
  {author} {\bibfnamefont {K.~M.}\ \bibnamefont {Svore}}, \bibinfo {author}
  {\bibfnamefont {T.}~\bibnamefont {Hoefler}}, \bibinfo {author} {\bibfnamefont
  {V.}~\bibnamefont {Kliuchnikov}}, \bibinfo {author} {\bibfnamefont {G.~H.}\
  \bibnamefont {Low}}, \bibinfo {author} {\bibfnamefont {M.}~\bibnamefont
  {Soeken}}, \bibinfo {author} {\bibfnamefont {A.}~\bibnamefont {Sundaram}}, \
  and\ \bibinfo {author} {\bibfnamefont {A.}~\bibnamefont {Vaschillo}},\ }\href
  {\doibase 10.48550/arXiv.2211.07629} {\enquote {\bibinfo {title} {Assessing
  requirements to scale to practical quantum advantage},}\ } (\bibinfo {year}
  {2022}),\ \bibinfo {note} {arXiv:2211.07629 [quant-ph]}\BibitemShut {NoStop}%
\bibitem [{\citenamefont {Gidney}\ and\ \citenamefont
  {Ekerå}(2021)}]{gidney_how_2021}%
  \BibitemOpen
  \bibfield  {author} {\bibinfo {author} {\bibfnamefont {C.}~\bibnamefont
  {Gidney}}\ and\ \bibinfo {author} {\bibfnamefont {M.}~\bibnamefont
  {Ekerå}},\ }\href {\doibase 10.22331/Q-2021-04-15-433} {\bibfield  {journal}
  {\bibinfo  {journal} {Quantum}\ }\textbf {\bibinfo {volume} {5}},\ \bibinfo
  {pages} {1} (\bibinfo {year} {2021})},\ \bibinfo {note} {arXiv: 1905.09749
  Publisher: Verein zur Forderung des Open Access Publizierens in den
  Quantenwissenschaften}\BibitemShut {NoStop}%
\bibitem [{\citenamefont {Loss}\ and\ \citenamefont
  {DiVincenzo}(1998)}]{loss_quantum_1998}%
  \BibitemOpen
  \bibfield  {author} {\bibinfo {author} {\bibfnamefont {D.}~\bibnamefont
  {Loss}}\ and\ \bibinfo {author} {\bibfnamefont {D.~P.}\ \bibnamefont
  {DiVincenzo}},\ }\href {\doibase 10.1103/PhysRevA.57.120} {\bibfield
  {journal} {\bibinfo  {journal} {Physical Review A - Atomic, Molecular, and
  Optical Physics}\ }\textbf {\bibinfo {volume} {57}},\ \bibinfo {pages} {120}
  (\bibinfo {year} {1998})},\ \bibinfo {note} {arXiv: cond-mat/9701055
  Publisher: American Physical Society}\BibitemShut {NoStop}%
\bibitem [{\citenamefont {Li}\ \emph {et~al.}(2010)\citenamefont {Li},
  \citenamefont {Cywinski}, \citenamefont {Culcer}, \citenamefont {Hu},\ and\
  \citenamefont {Das~Sarma}}]{Li2010}%
  \BibitemOpen
  \bibfield  {author} {\bibinfo {author} {\bibfnamefont {Q.}~\bibnamefont
  {Li}}, \bibinfo {author} {\bibfnamefont {L.}~\bibnamefont {Cywinski}},
  \bibinfo {author} {\bibfnamefont {D.}~\bibnamefont {Culcer}}, \bibinfo
  {author} {\bibfnamefont {X.}~\bibnamefont {Hu}}, \ and\ \bibinfo {author}
  {\bibfnamefont {S.}~\bibnamefont {Das~Sarma}},\ }\href {\doibase
  10.1103/PhysRevB.81.085313} {\bibfield  {journal} {\bibinfo  {journal}
  {Physical Review B - Condensed Matter and Materials Physics}\ }\textbf
  {\bibinfo {volume} {81}},\ \bibinfo {pages} {085313} (\bibinfo {year}
  {2010})},\ \bibinfo {note} {arXiv: 0906.4793 Publisher: American Physical
  Society}\BibitemShut {NoStop}%
\bibitem [{\citenamefont {Burkard}\ \emph {et~al.}(2021)\citenamefont
  {Burkard}, \citenamefont {Ladd}, \citenamefont {Nichol}, \citenamefont
  {Pan},\ and\ \citenamefont {Petta}}]{burkard_semiconductor_2021}%
  \BibitemOpen
  \bibfield  {author} {\bibinfo {author} {\bibfnamefont {G.}~\bibnamefont
  {Burkard}}, \bibinfo {author} {\bibfnamefont {T.~D.}\ \bibnamefont {Ladd}},
  \bibinfo {author} {\bibfnamefont {J.~M.}\ \bibnamefont {Nichol}}, \bibinfo
  {author} {\bibfnamefont {A.}~\bibnamefont {Pan}}, \ and\ \bibinfo {author}
  {\bibfnamefont {J.~R.}\ \bibnamefont {Petta}},\ }\href {\doibase
  10.48550/arXiv.2112.08863} {\enquote {\bibinfo {title} {Semiconductor {Spin}
  {Qubits}},}\ } (\bibinfo {year} {2021}),\ \bibinfo {note} {arXiv:2112.08863
  [cond-mat, physics:physics, physics:quant-ph]}\BibitemShut {NoStop}%
\bibitem [{\citenamefont {Hendrickx}\ \emph {et~al.}(2021)\citenamefont
  {Hendrickx}, \citenamefont {Lawrie}, \citenamefont {Russ}, \citenamefont {van
  Riggelen}, \citenamefont {de~Snoo}, \citenamefont {Schouten}, \citenamefont
  {Sammak}, \citenamefont {Scappucci},\ and\ \citenamefont
  {Veldhorst}}]{hendrickx_four-qubit_2021}%
  \BibitemOpen
  \bibfield  {author} {\bibinfo {author} {\bibfnamefont {N.~W.}\ \bibnamefont
  {Hendrickx}}, \bibinfo {author} {\bibfnamefont {W.~I.~L.}\ \bibnamefont
  {Lawrie}}, \bibinfo {author} {\bibfnamefont {M.}~\bibnamefont {Russ}},
  \bibinfo {author} {\bibfnamefont {F.}~\bibnamefont {van Riggelen}}, \bibinfo
  {author} {\bibfnamefont {S.~L.}\ \bibnamefont {de~Snoo}}, \bibinfo {author}
  {\bibfnamefont {R.~N.}\ \bibnamefont {Schouten}}, \bibinfo {author}
  {\bibfnamefont {A.}~\bibnamefont {Sammak}}, \bibinfo {author} {\bibfnamefont
  {G.}~\bibnamefont {Scappucci}}, \ and\ \bibinfo {author} {\bibfnamefont
  {M.}~\bibnamefont {Veldhorst}},\ }\href {\doibase 10.1038/s41586-021-03332-6}
  {\bibfield  {journal} {\bibinfo  {journal} {Nature}\ }\textbf {\bibinfo
  {volume} {591}},\ \bibinfo {pages} {580} (\bibinfo {year} {2021})},\ \bibinfo
  {note} {number: 7851 Publisher: Nature Publishing Group}\BibitemShut
  {NoStop}%
\bibitem [{\citenamefont {Hendrickx}\ \emph {et~al.}(2020)\citenamefont
  {Hendrickx}, \citenamefont {Franke}, \citenamefont {Sammak}, \citenamefont
  {Scappucci},\ and\ \citenamefont {Veldhorst}}]{hendrickx_fast_2020}%
  \BibitemOpen
  \bibfield  {author} {\bibinfo {author} {\bibfnamefont {N.~W.}\ \bibnamefont
  {Hendrickx}}, \bibinfo {author} {\bibfnamefont {D.~P.}\ \bibnamefont
  {Franke}}, \bibinfo {author} {\bibfnamefont {A.}~\bibnamefont {Sammak}},
  \bibinfo {author} {\bibfnamefont {G.}~\bibnamefont {Scappucci}}, \ and\
  \bibinfo {author} {\bibfnamefont {M.}~\bibnamefont {Veldhorst}},\ }\href
  {\doibase 10.1038/s41586-019-1919-3} {\bibfield  {journal} {\bibinfo
  {journal} {Nature}\ }\textbf {\bibinfo {volume} {577}},\ \bibinfo {pages}
  {487} (\bibinfo {year} {2020})},\ \bibinfo {note} {number: 7791 Publisher:
  Nature Publishing Group}\BibitemShut {NoStop}%
\bibitem [{\citenamefont {Zajac}\ \emph {et~al.}(2018)\citenamefont {Zajac},
  \citenamefont {Sigillito}, \citenamefont {Russ}, \citenamefont {Borjans},
  \citenamefont {Taylor}, \citenamefont {Burkard},\ and\ \citenamefont
  {Petta}}]{zajac_resonantly_2018}%
  \BibitemOpen
  \bibfield  {author} {\bibinfo {author} {\bibfnamefont {D.~M.}\ \bibnamefont
  {Zajac}}, \bibinfo {author} {\bibfnamefont {A.~J.}\ \bibnamefont
  {Sigillito}}, \bibinfo {author} {\bibfnamefont {M.}~\bibnamefont {Russ}},
  \bibinfo {author} {\bibfnamefont {F.}~\bibnamefont {Borjans}}, \bibinfo
  {author} {\bibfnamefont {J.~M.}\ \bibnamefont {Taylor}}, \bibinfo {author}
  {\bibfnamefont {G.}~\bibnamefont {Burkard}}, \ and\ \bibinfo {author}
  {\bibfnamefont {J.~R.}\ \bibnamefont {Petta}},\ }\href {\doibase
  10.1126/science.aao5965} {\bibfield  {journal} {\bibinfo  {journal}
  {Science}\ }\textbf {\bibinfo {volume} {359}},\ \bibinfo {pages} {439}
  (\bibinfo {year} {2018})},\ \bibinfo {note} {publisher: American Association
  for the Advancement of Science}\BibitemShut {NoStop}%
\bibitem [{\citenamefont {Xue}\ \emph {et~al.}(2022)\citenamefont {Xue},
  \citenamefont {Russ}, \citenamefont {Samkharadze}, \citenamefont {Undseth},
  \citenamefont {Sammak}, \citenamefont {Scappucci},\ and\ \citenamefont
  {Vandersypen}}]{xue_quantum_2022}%
  \BibitemOpen
  \bibfield  {author} {\bibinfo {author} {\bibfnamefont {X.}~\bibnamefont
  {Xue}}, \bibinfo {author} {\bibfnamefont {M.}~\bibnamefont {Russ}}, \bibinfo
  {author} {\bibfnamefont {N.}~\bibnamefont {Samkharadze}}, \bibinfo {author}
  {\bibfnamefont {B.}~\bibnamefont {Undseth}}, \bibinfo {author} {\bibfnamefont
  {A.}~\bibnamefont {Sammak}}, \bibinfo {author} {\bibfnamefont
  {G.}~\bibnamefont {Scappucci}}, \ and\ \bibinfo {author} {\bibfnamefont
  {L.~M.~K.}\ \bibnamefont {Vandersypen}},\ }\href {\doibase
  10.1038/s41586-021-04273-w} {\bibfield  {journal} {\bibinfo  {journal}
  {Nature}\ }\textbf {\bibinfo {volume} {601}},\ \bibinfo {pages} {343}
  (\bibinfo {year} {2022})},\ \bibinfo {note} {arXiv:2107.00628 [cond-mat,
  physics:quant-ph]}\BibitemShut {NoStop}%
\bibitem [{\citenamefont {Mills}\ \emph {et~al.}(2022)\citenamefont {Mills},
  \citenamefont {Guinn}, \citenamefont {Gullans}, \citenamefont {Sigillito},
  \citenamefont {Feldman}, \citenamefont {Nielsen},\ and\ \citenamefont
  {Petta}}]{mills_two-qubit_2022}%
  \BibitemOpen
  \bibfield  {author} {\bibinfo {author} {\bibfnamefont {A.~R.}\ \bibnamefont
  {Mills}}, \bibinfo {author} {\bibfnamefont {C.~R.}\ \bibnamefont {Guinn}},
  \bibinfo {author} {\bibfnamefont {M.~J.}\ \bibnamefont {Gullans}}, \bibinfo
  {author} {\bibfnamefont {A.~J.}\ \bibnamefont {Sigillito}}, \bibinfo {author}
  {\bibfnamefont {M.~M.}\ \bibnamefont {Feldman}}, \bibinfo {author}
  {\bibfnamefont {E.}~\bibnamefont {Nielsen}}, \ and\ \bibinfo {author}
  {\bibfnamefont {J.~R.}\ \bibnamefont {Petta}},\ }\href {\doibase
  10.1126/sciadv.abn5130} {\bibfield  {journal} {\bibinfo  {journal} {Science
  Advances}\ }\textbf {\bibinfo {volume} {8}},\ \bibinfo {pages} {eabn5130}
  (\bibinfo {year} {2022})},\ \bibinfo {note} {publisher: American Association
  for the Advancement of Science}\BibitemShut {NoStop}%
\bibitem [{\citenamefont {Huang}\ \emph {et~al.}(2019)\citenamefont {Huang},
  \citenamefont {Yang}, \citenamefont {Chan}, \citenamefont {Tanttu},
  \citenamefont {Hensen}, \citenamefont {Leon}, \citenamefont {Fogarty},
  \citenamefont {Hwang}, \citenamefont {Hudson}, \citenamefont {Itoh},
  \citenamefont {Morello}, \citenamefont {Laucht},\ and\ \citenamefont
  {Dzurak}}]{huang_fidelity_2019}%
  \BibitemOpen
  \bibfield  {author} {\bibinfo {author} {\bibfnamefont {W.}~\bibnamefont
  {Huang}}, \bibinfo {author} {\bibfnamefont {C.~H.}\ \bibnamefont {Yang}},
  \bibinfo {author} {\bibfnamefont {K.~W.}\ \bibnamefont {Chan}}, \bibinfo
  {author} {\bibfnamefont {T.}~\bibnamefont {Tanttu}}, \bibinfo {author}
  {\bibfnamefont {B.}~\bibnamefont {Hensen}}, \bibinfo {author} {\bibfnamefont
  {R.~C.~C.}\ \bibnamefont {Leon}}, \bibinfo {author} {\bibfnamefont {M.~A.}\
  \bibnamefont {Fogarty}}, \bibinfo {author} {\bibfnamefont {J.~C.~C.}\
  \bibnamefont {Hwang}}, \bibinfo {author} {\bibfnamefont {F.~E.}\ \bibnamefont
  {Hudson}}, \bibinfo {author} {\bibfnamefont {K.~M.}\ \bibnamefont {Itoh}},
  \bibinfo {author} {\bibfnamefont {A.}~\bibnamefont {Morello}}, \bibinfo
  {author} {\bibfnamefont {A.}~\bibnamefont {Laucht}}, \ and\ \bibinfo {author}
  {\bibfnamefont {A.~S.}\ \bibnamefont {Dzurak}},\ }\href {\doibase
  10.1038/s41586-019-1197-0} {\bibfield  {journal} {\bibinfo  {journal}
  {Nature}\ }\textbf {\bibinfo {volume} {569}},\ \bibinfo {pages} {532}
  (\bibinfo {year} {2019})},\ \bibinfo {note} {number: 7757 Publisher: Nature
  Publishing Group}\BibitemShut {NoStop}%
\bibitem [{\citenamefont {Tanttu}\ \emph {et~al.}(2023)\citenamefont {Tanttu},
  \citenamefont {Lim}, \citenamefont {Huang}, \citenamefont {Stuyck},
  \citenamefont {Gilbert}, \citenamefont {Su}, \citenamefont {Feng},
  \citenamefont {Cifuentes}, \citenamefont {Seedhouse}, \citenamefont
  {Seritan}, \citenamefont {Ostrove}, \citenamefont {Rudinger}, \citenamefont
  {Leon}, \citenamefont {Huang}, \citenamefont {Escott}, \citenamefont {Itoh},
  \citenamefont {Abrosimov}, \citenamefont {Pohl}, \citenamefont {Thewalt},
  \citenamefont {Hudson}, \citenamefont {Blume-Kohout}, \citenamefont
  {Bartlett}, \citenamefont {Morello}, \citenamefont {Laucht}, \citenamefont
  {Yang}, \citenamefont {Saraiva},\ and\ \citenamefont
  {Dzurak}}]{tanttu2023stability}%
  \BibitemOpen
  \bibfield  {author} {\bibinfo {author} {\bibfnamefont {T.}~\bibnamefont
  {Tanttu}}, \bibinfo {author} {\bibfnamefont {W.~H.}\ \bibnamefont {Lim}},
  \bibinfo {author} {\bibfnamefont {J.~Y.}\ \bibnamefont {Huang}}, \bibinfo
  {author} {\bibfnamefont {N.~D.}\ \bibnamefont {Stuyck}}, \bibinfo {author}
  {\bibfnamefont {W.}~\bibnamefont {Gilbert}}, \bibinfo {author} {\bibfnamefont
  {R.~Y.}\ \bibnamefont {Su}}, \bibinfo {author} {\bibfnamefont
  {M.}~\bibnamefont {Feng}}, \bibinfo {author} {\bibfnamefont {J.~D.}\
  \bibnamefont {Cifuentes}}, \bibinfo {author} {\bibfnamefont {A.~E.}\
  \bibnamefont {Seedhouse}}, \bibinfo {author} {\bibfnamefont {S.~K.}\
  \bibnamefont {Seritan}}, \bibinfo {author} {\bibfnamefont {C.~I.}\
  \bibnamefont {Ostrove}}, \bibinfo {author} {\bibfnamefont {K.~M.}\
  \bibnamefont {Rudinger}}, \bibinfo {author} {\bibfnamefont {R.~C.~C.}\
  \bibnamefont {Leon}}, \bibinfo {author} {\bibfnamefont {W.}~\bibnamefont
  {Huang}}, \bibinfo {author} {\bibfnamefont {C.~C.}\ \bibnamefont {Escott}},
  \bibinfo {author} {\bibfnamefont {K.~M.}\ \bibnamefont {Itoh}}, \bibinfo
  {author} {\bibfnamefont {N.~V.}\ \bibnamefont {Abrosimov}}, \bibinfo {author}
  {\bibfnamefont {H.-J.}\ \bibnamefont {Pohl}}, \bibinfo {author}
  {\bibfnamefont {M.~L.~W.}\ \bibnamefont {Thewalt}}, \bibinfo {author}
  {\bibfnamefont {F.~E.}\ \bibnamefont {Hudson}}, \bibinfo {author}
  {\bibfnamefont {R.}~\bibnamefont {Blume-Kohout}}, \bibinfo {author}
  {\bibfnamefont {S.~D.}\ \bibnamefont {Bartlett}}, \bibinfo {author}
  {\bibfnamefont {A.}~\bibnamefont {Morello}}, \bibinfo {author} {\bibfnamefont
  {A.}~\bibnamefont {Laucht}}, \bibinfo {author} {\bibfnamefont {C.~H.}\
  \bibnamefont {Yang}}, \bibinfo {author} {\bibfnamefont {A.}~\bibnamefont
  {Saraiva}}, \ and\ \bibinfo {author} {\bibfnamefont {A.~S.}\ \bibnamefont
  {Dzurak}},\ }\href@noop {} {\enquote {\bibinfo {title} {Stability of
  high-fidelity two-qubit operations in silicon},}\ } (\bibinfo {year}
  {2023}),\ \Eprint {http://arxiv.org/abs/2303.04090} {arXiv:2303.04090
  [quant-ph]} \BibitemShut {NoStop}%
\bibitem [{\citenamefont {Saraiva}\ \emph {et~al.}(2007)\citenamefont
  {Saraiva}, \citenamefont {Calderon},\ and\ \citenamefont
  {Koiller}}]{Saraiva2007}%
  \BibitemOpen
  \bibfield  {author} {\bibinfo {author} {\bibfnamefont {A.~L.}\ \bibnamefont
  {Saraiva}}, \bibinfo {author} {\bibfnamefont {M.~J.}\ \bibnamefont
  {Calderon}}, \ and\ \bibinfo {author} {\bibfnamefont {B.}~\bibnamefont
  {Koiller}},\ }\href {\doibase 10.1103/PhysRevB.76.233302} {\bibfield
  {journal} {\bibinfo  {journal} {Physical Review B - Condensed Matter and
  Materials Physics}\ }\textbf {\bibinfo {volume} {76}},\ \bibinfo {pages}
  {233302} (\bibinfo {year} {2007})},\ \bibinfo {note} {publisher: American
  Physical Society}\BibitemShut {NoStop}%
\bibitem [{\citenamefont {Koiller}\ \emph {et~al.}(2001)\citenamefont
  {Koiller}, \citenamefont {Hu},\ and\ \citenamefont
  {Das~Sarma}}]{Koiller2002}%
  \BibitemOpen
  \bibfield  {author} {\bibinfo {author} {\bibfnamefont {B.}~\bibnamefont
  {Koiller}}, \bibinfo {author} {\bibfnamefont {X.}~\bibnamefont {Hu}}, \ and\
  \bibinfo {author} {\bibfnamefont {S.}~\bibnamefont {Das~Sarma}},\ }\href
  {\doibase 10.1103/PhysRevLett.88.027903} {\bibfield  {journal} {\bibinfo
  {journal} {Phys. Rev. Lett.}\ }\textbf {\bibinfo {volume} {88}},\ \bibinfo
  {pages} {027903} (\bibinfo {year} {2001})}\BibitemShut {NoStop}%
\bibitem [{\citenamefont {Saraiva}\ \emph {et~al.}(2022)\citenamefont
  {Saraiva}, \citenamefont {Lim}, \citenamefont {Yang}, \citenamefont {Escott},
  \citenamefont {Laucht},\ and\ \citenamefont {Dzurak}}]{Saraiva2022}%
  \BibitemOpen
  \bibfield  {author} {\bibinfo {author} {\bibfnamefont {A.}~\bibnamefont
  {Saraiva}}, \bibinfo {author} {\bibfnamefont {W.~H.}\ \bibnamefont {Lim}},
  \bibinfo {author} {\bibfnamefont {C.~H.}\ \bibnamefont {Yang}}, \bibinfo
  {author} {\bibfnamefont {C.~C.}\ \bibnamefont {Escott}}, \bibinfo {author}
  {\bibfnamefont {A.}~\bibnamefont {Laucht}}, \ and\ \bibinfo {author}
  {\bibfnamefont {A.~S.}\ \bibnamefont {Dzurak}},\ }\href {\doibase
  10.1002/adfm.202105488} {\bibfield  {journal} {\bibinfo  {journal} {Advanced
  Functional Materials}\ }\textbf {\bibinfo {volume} {32}},\ \bibinfo {pages}
  {2105488} (\bibinfo {year} {2022})},\ \bibinfo {note} {publisher: John Wiley
  and Sons Inc}\BibitemShut {NoStop}%
\bibitem [{\citenamefont {Geyer}\ \emph {et~al.}(2021)\citenamefont {Geyer},
  \citenamefont {Camenzind}, \citenamefont {Czornomaz}, \citenamefont
  {Deshpande}, \citenamefont {Fuhrer}, \citenamefont {Warburton}, \citenamefont
  {Zumbühl},\ and\ \citenamefont {Kuhlmann}}]{geyer_silicon_2021}%
  \BibitemOpen
  \bibfield  {author} {\bibinfo {author} {\bibfnamefont {S.}~\bibnamefont
  {Geyer}}, \bibinfo {author} {\bibfnamefont {L.~C.}\ \bibnamefont
  {Camenzind}}, \bibinfo {author} {\bibfnamefont {L.}~\bibnamefont
  {Czornomaz}}, \bibinfo {author} {\bibfnamefont {V.}~\bibnamefont
  {Deshpande}}, \bibinfo {author} {\bibfnamefont {A.}~\bibnamefont {Fuhrer}},
  \bibinfo {author} {\bibfnamefont {R.~J.}\ \bibnamefont {Warburton}}, \bibinfo
  {author} {\bibfnamefont {D.~M.}\ \bibnamefont {Zumbühl}}, \ and\ \bibinfo
  {author} {\bibfnamefont {A.~V.}\ \bibnamefont {Kuhlmann}},\ }\href {\doibase
  10.1063/5.0036520} {\bibfield  {journal} {\bibinfo  {journal} {Applied
  Physics Letters}\ }\textbf {\bibinfo {volume} {118}},\ \bibinfo {pages}
  {104004} (\bibinfo {year} {2021})},\ \bibinfo {note} {arXiv:2007.15400
  [cond-mat]}\BibitemShut {NoStop}%
\bibitem [{\citenamefont {Petit}\ \emph
  {et~al.}(2020{\natexlab{a}})\citenamefont {Petit}, \citenamefont {Eenink},
  \citenamefont {Russ}, \citenamefont {Lawrie}, \citenamefont {Hendrickx},
  \citenamefont {Philips}, \citenamefont {Clarke}, \citenamefont
  {Vandersypen},\ and\ \citenamefont {Veldhorst}}]{petit_universal_2020}%
  \BibitemOpen
  \bibfield  {author} {\bibinfo {author} {\bibfnamefont {L.}~\bibnamefont
  {Petit}}, \bibinfo {author} {\bibfnamefont {H.~G.~J.}\ \bibnamefont
  {Eenink}}, \bibinfo {author} {\bibfnamefont {M.}~\bibnamefont {Russ}},
  \bibinfo {author} {\bibfnamefont {W.~I.~L.}\ \bibnamefont {Lawrie}}, \bibinfo
  {author} {\bibfnamefont {N.~W.}\ \bibnamefont {Hendrickx}}, \bibinfo {author}
  {\bibfnamefont {S.~G.~J.}\ \bibnamefont {Philips}}, \bibinfo {author}
  {\bibfnamefont {J.~S.}\ \bibnamefont {Clarke}}, \bibinfo {author}
  {\bibfnamefont {L.~M.~K.}\ \bibnamefont {Vandersypen}}, \ and\ \bibinfo
  {author} {\bibfnamefont {M.}~\bibnamefont {Veldhorst}},\ }\href {\doibase
  10.1038/s41586-020-2170-7} {\bibfield  {journal} {\bibinfo  {journal}
  {Nature}\ }\textbf {\bibinfo {volume} {580}},\ \bibinfo {pages} {355}
  (\bibinfo {year} {2020}{\natexlab{a}})},\ \bibinfo {note} {number: 7803
  Publisher: Nature Publishing Group}\BibitemShut {NoStop}%
\bibitem [{\citenamefont {Petit}\ \emph
  {et~al.}(2020{\natexlab{b}})\citenamefont {Petit}, \citenamefont {Russ},
  \citenamefont {Eenink}, \citenamefont {Lawrie}, \citenamefont {Clarke},
  \citenamefont {Vandersypen},\ and\ \citenamefont
  {Veldhorst}}]{petit_high-fidelity_2020}%
  \BibitemOpen
  \bibfield  {author} {\bibinfo {author} {\bibfnamefont {L.}~\bibnamefont
  {Petit}}, \bibinfo {author} {\bibfnamefont {M.}~\bibnamefont {Russ}},
  \bibinfo {author} {\bibfnamefont {H.~G.~J.}\ \bibnamefont {Eenink}}, \bibinfo
  {author} {\bibfnamefont {W.~I.~L.}\ \bibnamefont {Lawrie}}, \bibinfo {author}
  {\bibfnamefont {J.~S.}\ \bibnamefont {Clarke}}, \bibinfo {author}
  {\bibfnamefont {L.~M.~K.}\ \bibnamefont {Vandersypen}}, \ and\ \bibinfo
  {author} {\bibfnamefont {M.}~\bibnamefont {Veldhorst}},\ }\href {\doibase
  10.48550/arXiv.2007.09034} {\enquote {\bibinfo {title} {High-fidelity
  two-qubit gates in silicon above one {Kelvin}},}\ } (\bibinfo {year}
  {2020}{\natexlab{b}}),\ \bibinfo {note} {arXiv:2007.09034
  [cond-mat]}\BibitemShut {NoStop}%
\bibitem [{\citenamefont {Takeda}\ \emph {et~al.}(2022)\citenamefont {Takeda},
  \citenamefont {Noiri}, \citenamefont {Nakajima}, \citenamefont {Kobayashi},\
  and\ \citenamefont {Tarucha}}]{takeda_quantum_2022}%
  \BibitemOpen
  \bibfield  {author} {\bibinfo {author} {\bibfnamefont {K.}~\bibnamefont
  {Takeda}}, \bibinfo {author} {\bibfnamefont {A.}~\bibnamefont {Noiri}},
  \bibinfo {author} {\bibfnamefont {T.}~\bibnamefont {Nakajima}}, \bibinfo
  {author} {\bibfnamefont {T.}~\bibnamefont {Kobayashi}}, \ and\ \bibinfo
  {author} {\bibfnamefont {S.}~\bibnamefont {Tarucha}},\ }\href {\doibase
  10.1038/s41586-022-04986-6} {\bibfield  {journal} {\bibinfo  {journal}
  {Nature}\ }\textbf {\bibinfo {volume} {608}},\ \bibinfo {pages} {682}
  (\bibinfo {year} {2022})},\ \bibinfo {note} {number: 7924 Publisher: Nature
  Publishing Group}\BibitemShut {NoStop}%
\bibitem [{\citenamefont {Leon}\ \emph {et~al.}(2021)\citenamefont {Leon},
  \citenamefont {Yang}, \citenamefont {Hwang}, \citenamefont {Camirand~Lemyre},
  \citenamefont {Tanttu}, \citenamefont {Huang}, \citenamefont {Huang},
  \citenamefont {Hudson}, \citenamefont {Itoh}, \citenamefont {Laucht},
  \citenamefont {Pioro-Ladrière}, \citenamefont {Saraiva},\ and\ \citenamefont
  {Dzurak}}]{Leon2021}%
  \BibitemOpen
  \bibfield  {author} {\bibinfo {author} {\bibfnamefont {R.~C.}\ \bibnamefont
  {Leon}}, \bibinfo {author} {\bibfnamefont {C.~H.}\ \bibnamefont {Yang}},
  \bibinfo {author} {\bibfnamefont {J.~C.}\ \bibnamefont {Hwang}}, \bibinfo
  {author} {\bibfnamefont {J.}~\bibnamefont {Camirand~Lemyre}}, \bibinfo
  {author} {\bibfnamefont {T.}~\bibnamefont {Tanttu}}, \bibinfo {author}
  {\bibfnamefont {W.}~\bibnamefont {Huang}}, \bibinfo {author} {\bibfnamefont
  {J.~Y.}\ \bibnamefont {Huang}}, \bibinfo {author} {\bibfnamefont {F.~E.}\
  \bibnamefont {Hudson}}, \bibinfo {author} {\bibfnamefont {K.~M.}\
  \bibnamefont {Itoh}}, \bibinfo {author} {\bibfnamefont {A.}~\bibnamefont
  {Laucht}}, \bibinfo {author} {\bibfnamefont {M.}~\bibnamefont
  {Pioro-Ladrière}}, \bibinfo {author} {\bibfnamefont {A.}~\bibnamefont
  {Saraiva}}, \ and\ \bibinfo {author} {\bibfnamefont {A.~S.}\ \bibnamefont
  {Dzurak}},\ }\href {\doibase 10.1038/s41467-021-23437-w} {\bibfield
  {journal} {\bibinfo  {journal} {Nature Communications}\ }\textbf {\bibinfo
  {volume} {12}},\ \bibinfo {pages} {1} (\bibinfo {year} {2021})},\ \bibinfo
  {note} {arXiv: 2008.03968 Publisher: Nature Research}\BibitemShut {NoStop}%
\bibitem [{\citenamefont {Shehata}\ \emph {et~al.}(2022)\citenamefont
  {Shehata}, \citenamefont {Simion}, \citenamefont {Li}, \citenamefont
  {Mohiyaddin}, \citenamefont {Wan}, \citenamefont {Mongillo}, \citenamefont
  {Govoreanu}, \citenamefont {Radu}, \citenamefont {Greve},\ and\ \citenamefont
  {Dorpe}}]{shehata2022modelling}%
  \BibitemOpen
  \bibfield  {author} {\bibinfo {author} {\bibfnamefont {M.~M. E.~K.}\
  \bibnamefont {Shehata}}, \bibinfo {author} {\bibfnamefont {G.}~\bibnamefont
  {Simion}}, \bibinfo {author} {\bibfnamefont {R.}~\bibnamefont {Li}}, \bibinfo
  {author} {\bibfnamefont {F.~A.}\ \bibnamefont {Mohiyaddin}}, \bibinfo
  {author} {\bibfnamefont {D.}~\bibnamefont {Wan}}, \bibinfo {author}
  {\bibfnamefont {M.}~\bibnamefont {Mongillo}}, \bibinfo {author}
  {\bibfnamefont {B.}~\bibnamefont {Govoreanu}}, \bibinfo {author}
  {\bibfnamefont {I.}~\bibnamefont {Radu}}, \bibinfo {author} {\bibfnamefont
  {K.~D.}\ \bibnamefont {Greve}}, \ and\ \bibinfo {author} {\bibfnamefont
  {P.~V.}\ \bibnamefont {Dorpe}},\ }\href@noop {} {\enquote {\bibinfo {title}
  {Modelling semiconductor spin qubits and their charge noise environment for
  quantum gate fidelity estimation},}\ } (\bibinfo {year} {2022}),\ \Eprint
  {http://arxiv.org/abs/2210.04539} {arXiv:2210.04539 [cond-mat.mes-hall]}
  \BibitemShut {NoStop}%
\bibitem [{\citenamefont {Cifuentes}\ \emph {et~al.}(2023)\citenamefont
  {Cifuentes}, \citenamefont {Tanttu}, \citenamefont {Gilbert}, \citenamefont
  {Huang}, \citenamefont {Vahapoglu}, \citenamefont {Leon}, \citenamefont
  {Serrano}, \citenamefont {Otter}, \citenamefont {Dunmore}, \citenamefont
  {Mai}, \citenamefont {Schlattner}, \citenamefont {Feng}, \citenamefont
  {Itoh}, \citenamefont {Abrosimov}, \citenamefont {Pohl}, \citenamefont
  {Thewalt}, \citenamefont {Laucht}, \citenamefont {Yang}, \citenamefont
  {Escott}, \citenamefont {Lim}, \citenamefont {Hudson}, \citenamefont
  {Rahman}, \citenamefont {Saraiva},\ and\ \citenamefont
  {Dzurak}}]{cifuentes2023bounds}%
  \BibitemOpen
  \bibfield  {author} {\bibinfo {author} {\bibfnamefont {J.~D.}\ \bibnamefont
  {Cifuentes}}, \bibinfo {author} {\bibfnamefont {T.}~\bibnamefont {Tanttu}},
  \bibinfo {author} {\bibfnamefont {W.}~\bibnamefont {Gilbert}}, \bibinfo
  {author} {\bibfnamefont {J.~Y.}\ \bibnamefont {Huang}}, \bibinfo {author}
  {\bibfnamefont {E.}~\bibnamefont {Vahapoglu}}, \bibinfo {author}
  {\bibfnamefont {R.~C.~C.}\ \bibnamefont {Leon}}, \bibinfo {author}
  {\bibfnamefont {S.}~\bibnamefont {Serrano}}, \bibinfo {author} {\bibfnamefont
  {D.}~\bibnamefont {Otter}}, \bibinfo {author} {\bibfnamefont
  {D.}~\bibnamefont {Dunmore}}, \bibinfo {author} {\bibfnamefont {P.~Y.}\
  \bibnamefont {Mai}}, \bibinfo {author} {\bibfnamefont {F.}~\bibnamefont
  {Schlattner}}, \bibinfo {author} {\bibfnamefont {M.}~\bibnamefont {Feng}},
  \bibinfo {author} {\bibfnamefont {K.}~\bibnamefont {Itoh}}, \bibinfo {author}
  {\bibfnamefont {N.}~\bibnamefont {Abrosimov}}, \bibinfo {author}
  {\bibfnamefont {H.-J.}\ \bibnamefont {Pohl}}, \bibinfo {author}
  {\bibfnamefont {M.}~\bibnamefont {Thewalt}}, \bibinfo {author} {\bibfnamefont
  {A.}~\bibnamefont {Laucht}}, \bibinfo {author} {\bibfnamefont {C.~H.}\
  \bibnamefont {Yang}}, \bibinfo {author} {\bibfnamefont {C.~C.}\ \bibnamefont
  {Escott}}, \bibinfo {author} {\bibfnamefont {W.~H.}\ \bibnamefont {Lim}},
  \bibinfo {author} {\bibfnamefont {F.~E.}\ \bibnamefont {Hudson}}, \bibinfo
  {author} {\bibfnamefont {R.}~\bibnamefont {Rahman}}, \bibinfo {author}
  {\bibfnamefont {A.}~\bibnamefont {Saraiva}}, \ and\ \bibinfo {author}
  {\bibfnamefont {A.~S.}\ \bibnamefont {Dzurak}},\ }\href@noop {} {\enquote
  {\bibinfo {title} {Bounds to electron spin qubit variability for scalable
  cmos architectures},}\ } (\bibinfo {year} {2023}),\ \Eprint
  {http://arxiv.org/abs/2303.14864} {arXiv:2303.14864 [quant-ph]} \BibitemShut
  {NoStop}%
\bibitem [{\citenamefont {Ceperley}(1995)}]{Ceperley1995}%
  \BibitemOpen
  \bibfield  {author} {\bibinfo {author} {\bibfnamefont {D.~M.}\ \bibnamefont
  {Ceperley}},\ }\href {\doibase 10.1103/RevModPhys.67.279} {\bibfield
  {journal} {\bibinfo  {journal} {Reviews of Modern Physics}\ }\textbf
  {\bibinfo {volume} {67}},\ \bibinfo {pages} {279} (\bibinfo {year}
  {1995})}\BibitemShut {NoStop}%
\bibitem [{\citenamefont {Yan}\ and\ \citenamefont {Blume}(2017)}]{Yan_2017}%
  \BibitemOpen
  \bibfield  {author} {\bibinfo {author} {\bibfnamefont {Y.}~\bibnamefont
  {Yan}}\ and\ \bibinfo {author} {\bibfnamefont {D.}~\bibnamefont {Blume}},\
  }\href {\doibase 10.1088/1361-6455/aa8d7f} {\bibfield  {journal} {\bibinfo
  {journal} {Journal of Physics B: Atomic, Molecular and Optical Physics}\
  }\textbf {\bibinfo {volume} {50}},\ \bibinfo {pages} {223001} (\bibinfo
  {year} {2017})}\BibitemShut {NoStop}%
\bibitem [{\citenamefont {Niquet}\ \emph {et~al.}(2020)\citenamefont {Niquet},
  \citenamefont {Hutin}, \citenamefont {Diaz}, \citenamefont {Venitucci},
  \citenamefont {Li}, \citenamefont {Michal}, \citenamefont {Fernández-Bada},
  \citenamefont {Jacquinot}, \citenamefont {Amisse}, \citenamefont {Apra},
  \citenamefont {Ezzouch}, \citenamefont {Piot}, \citenamefont {Vincent},
  \citenamefont {Yu}, \citenamefont {Zihlmann}, \citenamefont {Brun-Barrière},
  \citenamefont {Schmitt}, \citenamefont {Dumur}, \citenamefont {Maurand},
  \citenamefont {Jehl}, \citenamefont {Sanquer}, \citenamefont {Bertrand},
  \citenamefont {Rambal}, \citenamefont {Niebojewski}, \citenamefont
  {Bedecarrats}, \citenamefont {Cassé}, \citenamefont {Catapano},
  \citenamefont {Mortemousque}, \citenamefont {Thomas}, \citenamefont
  {Thonnart}, \citenamefont {Billiot}, \citenamefont {Morel}, \citenamefont
  {Charbonnier}, \citenamefont {Pallegoix}, \citenamefont {Niegemann},
  \citenamefont {Klemt}, \citenamefont {Urdampilleta}, \citenamefont
  {El~Homsy}, \citenamefont {Nurizzo}, \citenamefont {Chanrion}, \citenamefont
  {Jadot}, \citenamefont {Spence}, \citenamefont {Thiney}, \citenamefont {Paz},
  \citenamefont {de~Franceschi}, \citenamefont {Vinet},\ and\ \citenamefont
  {Meunier}}]{Niquet_2020}%
  \BibitemOpen
  \bibfield  {author} {\bibinfo {author} {\bibfnamefont {Y.~M.}\ \bibnamefont
  {Niquet}}, \bibinfo {author} {\bibfnamefont {L.}~\bibnamefont {Hutin}},
  \bibinfo {author} {\bibfnamefont {B.~M.}\ \bibnamefont {Diaz}}, \bibinfo
  {author} {\bibfnamefont {B.}~\bibnamefont {Venitucci}}, \bibinfo {author}
  {\bibfnamefont {J.}~\bibnamefont {Li}}, \bibinfo {author} {\bibfnamefont
  {V.}~\bibnamefont {Michal}}, \bibinfo {author} {\bibfnamefont {G.~T.}\
  \bibnamefont {Fernández-Bada}}, \bibinfo {author} {\bibfnamefont
  {H.}~\bibnamefont {Jacquinot}}, \bibinfo {author} {\bibfnamefont
  {A.}~\bibnamefont {Amisse}}, \bibinfo {author} {\bibfnamefont
  {A.}~\bibnamefont {Apra}}, \bibinfo {author} {\bibfnamefont {R.}~\bibnamefont
  {Ezzouch}}, \bibinfo {author} {\bibfnamefont {N.}~\bibnamefont {Piot}},
  \bibinfo {author} {\bibfnamefont {E.}~\bibnamefont {Vincent}}, \bibinfo
  {author} {\bibfnamefont {C.}~\bibnamefont {Yu}}, \bibinfo {author}
  {\bibfnamefont {S.}~\bibnamefont {Zihlmann}}, \bibinfo {author}
  {\bibfnamefont {B.}~\bibnamefont {Brun-Barrière}}, \bibinfo {author}
  {\bibfnamefont {V.}~\bibnamefont {Schmitt}}, \bibinfo {author} {\bibfnamefont
  {E.}~\bibnamefont {Dumur}}, \bibinfo {author} {\bibfnamefont
  {R.}~\bibnamefont {Maurand}}, \bibinfo {author} {\bibfnamefont
  {X.}~\bibnamefont {Jehl}}, \bibinfo {author} {\bibfnamefont {M.}~\bibnamefont
  {Sanquer}}, \bibinfo {author} {\bibfnamefont {B.}~\bibnamefont {Bertrand}},
  \bibinfo {author} {\bibfnamefont {N.}~\bibnamefont {Rambal}}, \bibinfo
  {author} {\bibfnamefont {H.}~\bibnamefont {Niebojewski}}, \bibinfo {author}
  {\bibfnamefont {T.}~\bibnamefont {Bedecarrats}}, \bibinfo {author}
  {\bibfnamefont {M.}~\bibnamefont {Cassé}}, \bibinfo {author} {\bibfnamefont
  {E.}~\bibnamefont {Catapano}}, \bibinfo {author} {\bibfnamefont {P.~A.}\
  \bibnamefont {Mortemousque}}, \bibinfo {author} {\bibfnamefont
  {C.}~\bibnamefont {Thomas}}, \bibinfo {author} {\bibfnamefont
  {Y.}~\bibnamefont {Thonnart}}, \bibinfo {author} {\bibfnamefont
  {G.}~\bibnamefont {Billiot}}, \bibinfo {author} {\bibfnamefont
  {A.}~\bibnamefont {Morel}}, \bibinfo {author} {\bibfnamefont
  {J.}~\bibnamefont {Charbonnier}}, \bibinfo {author} {\bibfnamefont
  {L.}~\bibnamefont {Pallegoix}}, \bibinfo {author} {\bibfnamefont
  {D.}~\bibnamefont {Niegemann}}, \bibinfo {author} {\bibfnamefont
  {B.}~\bibnamefont {Klemt}}, \bibinfo {author} {\bibfnamefont
  {M.}~\bibnamefont {Urdampilleta}}, \bibinfo {author} {\bibfnamefont
  {V.}~\bibnamefont {El~Homsy}}, \bibinfo {author} {\bibfnamefont
  {M.}~\bibnamefont {Nurizzo}}, \bibinfo {author} {\bibfnamefont
  {E.}~\bibnamefont {Chanrion}}, \bibinfo {author} {\bibfnamefont
  {B.}~\bibnamefont {Jadot}}, \bibinfo {author} {\bibfnamefont
  {C.}~\bibnamefont {Spence}}, \bibinfo {author} {\bibfnamefont
  {V.}~\bibnamefont {Thiney}}, \bibinfo {author} {\bibfnamefont
  {B.}~\bibnamefont {Paz}}, \bibinfo {author} {\bibfnamefont {S.}~\bibnamefont
  {de~Franceschi}}, \bibinfo {author} {\bibfnamefont {M.}~\bibnamefont
  {Vinet}}, \ and\ \bibinfo {author} {\bibfnamefont {T.}~\bibnamefont
  {Meunier}},\ }in\ \href {\doibase 10.1109/IEDM13553.2020.9371962} {\emph
  {\bibinfo {booktitle} {2020 IEEE International Electron Devices Meeting
  (IEDM)}}}\ (\bibinfo {year} {2020})\ pp.\ \bibinfo {pages}
  {30.1.1--30.1.4}\BibitemShut {NoStop}%
\bibitem [{\citenamefont {Pedersen}\ \emph {et~al.}(2010)\citenamefont
  {Pedersen}, \citenamefont {Zhang}, \citenamefont {Gilbert},\ and\
  \citenamefont {Shumway}}]{Pedersen2010}%
  \BibitemOpen
  \bibfield  {author} {\bibinfo {author} {\bibfnamefont {J.~G.}\ \bibnamefont
  {Pedersen}}, \bibinfo {author} {\bibfnamefont {L.}~\bibnamefont {Zhang}},
  \bibinfo {author} {\bibfnamefont {M.~J.}\ \bibnamefont {Gilbert}}, \ and\
  \bibinfo {author} {\bibfnamefont {J.}~\bibnamefont {Shumway}},\ }\href
  {\doibase 10.1088/0953-8984/22/14/145301} {\bibfield  {journal} {\bibinfo
  {journal} {Journal of Physics Condensed Matter}\ }\textbf {\bibinfo {volume}
  {22}} (\bibinfo {year} {2010}),\ 10.1088/0953-8984/22/14/145301}\BibitemShut
  {NoStop}%
\bibitem [{\citenamefont {Martinez}\ and\ \citenamefont
  {Niquet}(2022)}]{Martinez_Variability_2022}%
  \BibitemOpen
  \bibfield  {author} {\bibinfo {author} {\bibfnamefont {B.}~\bibnamefont
  {Martinez}}\ and\ \bibinfo {author} {\bibfnamefont {Y.-M.}\ \bibnamefont
  {Niquet}},\ }\href {\doibase 10.1103/PhysRevApplied.17.024022} {\bibfield
  {journal} {\bibinfo  {journal} {Phys. Rev. Appl.}\ }\textbf {\bibinfo
  {volume} {17}},\ \bibinfo {pages} {024022} (\bibinfo {year}
  {2022})}\BibitemShut {NoStop}%
\bibitem [{\citenamefont {Zwanenburg}\ \emph {et~al.}(2013)\citenamefont
  {Zwanenburg}, \citenamefont {Dzurak}, \citenamefont {Morello}, \citenamefont
  {Simmons}, \citenamefont {Hollenberg}, \citenamefont {Klimeck}, \citenamefont
  {Rogge}, \citenamefont {Coppersmith},\ and\ \citenamefont
  {Eriksson}}]{Zwanenburg2013}%
  \BibitemOpen
  \bibfield  {author} {\bibinfo {author} {\bibfnamefont {F.~A.}\ \bibnamefont
  {Zwanenburg}}, \bibinfo {author} {\bibfnamefont {A.~S.}\ \bibnamefont
  {Dzurak}}, \bibinfo {author} {\bibfnamefont {A.}~\bibnamefont {Morello}},
  \bibinfo {author} {\bibfnamefont {M.~Y.}\ \bibnamefont {Simmons}}, \bibinfo
  {author} {\bibfnamefont {L.~C.~L.}\ \bibnamefont {Hollenberg}}, \bibinfo
  {author} {\bibfnamefont {G.}~\bibnamefont {Klimeck}}, \bibinfo {author}
  {\bibfnamefont {S.}~\bibnamefont {Rogge}}, \bibinfo {author} {\bibfnamefont
  {S.~N.}\ \bibnamefont {Coppersmith}}, \ and\ \bibinfo {author} {\bibfnamefont
  {M.~A.}\ \bibnamefont {Eriksson}},\ }\href {\doibase
  10.1103/RevModPhys.85.961} {\bibfield  {journal} {\bibinfo  {journal} {Rev.
  Mod. Phys.}\ }\textbf {\bibinfo {volume} {85}},\ \bibinfo {pages} {961}
  (\bibinfo {year} {2013})}\BibitemShut {NoStop}%
\bibitem [{\citenamefont {Saraiva}\ \emph {et~al.}(2009)\citenamefont
  {Saraiva}, \citenamefont {Calder\'on}, \citenamefont {Hu}, \citenamefont
  {Das~Sarma},\ and\ \citenamefont {Koiller}}]{Saraiva2009}%
  \BibitemOpen
  \bibfield  {author} {\bibinfo {author} {\bibfnamefont {A.~L.}\ \bibnamefont
  {Saraiva}}, \bibinfo {author} {\bibfnamefont {M.~J.}\ \bibnamefont
  {Calder\'on}}, \bibinfo {author} {\bibfnamefont {X.}~\bibnamefont {Hu}},
  \bibinfo {author} {\bibfnamefont {S.}~\bibnamefont {Das~Sarma}}, \ and\
  \bibinfo {author} {\bibfnamefont {B.}~\bibnamefont {Koiller}},\ }\href
  {\doibase 10.1103/PhysRevB.80.081305} {\bibfield  {journal} {\bibinfo
  {journal} {Phys. Rev. B}\ }\textbf {\bibinfo {volume} {80}},\ \bibinfo
  {pages} {081305(R)} (\bibinfo {year} {2009})}\BibitemShut {NoStop}%
\bibitem [{\citenamefont {Tariq}\ and\ \citenamefont
  {Hu}(2022)}]{tariq_impact_2022}%
  \BibitemOpen
  \bibfield  {author} {\bibinfo {author} {\bibfnamefont {B.}~\bibnamefont
  {Tariq}}\ and\ \bibinfo {author} {\bibfnamefont {X.}~\bibnamefont {Hu}},\
  }\href {\doibase 10.1038/s41534-022-00554-y} {\bibfield  {journal} {\bibinfo
  {journal} {npj Quantum Information}\ }\textbf {\bibinfo {volume} {8}},\
  \bibinfo {pages} {1} (\bibinfo {year} {2022})},\ \bibinfo {note} {number: 1
  Publisher: Nature Publishing Group}\BibitemShut {NoStop}%
\bibitem [{\citenamefont {Wang}\ \emph {et~al.}(2016)\citenamefont {Wang},
  \citenamefont {Tankasala}, \citenamefont {Hollenberg}, \citenamefont
  {Klimeck}, \citenamefont {Simmons},\ and\ \citenamefont
  {Rahman}}]{wang_highly_2016}%
  \BibitemOpen
  \bibfield  {author} {\bibinfo {author} {\bibfnamefont {Y.}~\bibnamefont
  {Wang}}, \bibinfo {author} {\bibfnamefont {A.}~\bibnamefont {Tankasala}},
  \bibinfo {author} {\bibfnamefont {L.~C.~L.}\ \bibnamefont {Hollenberg}},
  \bibinfo {author} {\bibfnamefont {G.}~\bibnamefont {Klimeck}}, \bibinfo
  {author} {\bibfnamefont {M.~Y.}\ \bibnamefont {Simmons}}, \ and\ \bibinfo
  {author} {\bibfnamefont {R.}~\bibnamefont {Rahman}},\ }\href {\doibase
  10.1038/npjqi.2016.8} {\bibfield  {journal} {\bibinfo  {journal} {npj Quantum
  Information}\ }\textbf {\bibinfo {volume} {2}},\ \bibinfo {pages} {1}
  (\bibinfo {year} {2016})},\ \bibinfo {note} {number: 1 Publisher: Nature
  Publishing Group}\BibitemShut {NoStop}%
\bibitem [{\citenamefont {Calderon}\ \emph {et~al.}(2007)\citenamefont
  {Calderon}, \citenamefont {Koiller},\ and\ \citenamefont
  {Das~Sarma}}]{Calderon_donor_2007}%
  \BibitemOpen
  \bibfield  {author} {\bibinfo {author} {\bibfnamefont {M.~J.}\ \bibnamefont
  {Calderon}}, \bibinfo {author} {\bibfnamefont {B.}~\bibnamefont {Koiller}}, \
  and\ \bibinfo {author} {\bibfnamefont {S.}~\bibnamefont {Das~Sarma}},\ }\href
  {\doibase 10.1103/PhysRevB.75.125311} {\bibfield  {journal} {\bibinfo
  {journal} {Phys. Rev. B}\ }\textbf {\bibinfo {volume} {75}},\ \bibinfo
  {pages} {125311} (\bibinfo {year} {2007})}\BibitemShut {NoStop}%
\bibitem [{\citenamefont {Westbroek}\ \emph {et~al.}(2018)\citenamefont
  {Westbroek}, \citenamefont {King}, \citenamefont {Vvedensky},\ and\
  \citenamefont {Dürr}}]{Westbroek_2018}%
  \BibitemOpen
  \bibfield  {author} {\bibinfo {author} {\bibfnamefont {M.~J.~E.}\
  \bibnamefont {Westbroek}}, \bibinfo {author} {\bibfnamefont {P.~R.}\
  \bibnamefont {King}}, \bibinfo {author} {\bibfnamefont {D.~D.}\ \bibnamefont
  {Vvedensky}}, \ and\ \bibinfo {author} {\bibfnamefont {S.}~\bibnamefont
  {Dürr}},\ }\href {\doibase 10.1119/1.5024926} {\bibfield  {journal}
  {\bibinfo  {journal} {American Journal of Physics}\ }\textbf {\bibinfo
  {volume} {86}},\ \bibinfo {pages} {293} (\bibinfo {year} {2018})}\BibitemShut
  {NoStop}%
\bibitem [{\citenamefont {Dornheim}(2019)}]{Dornheim2019}%
  \BibitemOpen
  \bibfield  {author} {\bibinfo {author} {\bibfnamefont {T.}~\bibnamefont
  {Dornheim}},\ }\href {\doibase 10.1103/PhysRevE.100.023307} {\bibfield
  {journal} {\bibinfo  {journal} {Physical Review E}\ }\textbf {\bibinfo
  {volume} {100}},\ \bibinfo {pages} {023307} (\bibinfo {year} {2019})},\
  \bibinfo {note} {publisher: American Physical Society}\BibitemShut {NoStop}%
\bibitem [{\citenamefont {Weiss}\ and\ \citenamefont
  {Egger}(2005)}]{Weiss2005}%
  \BibitemOpen
  \bibfield  {author} {\bibinfo {author} {\bibfnamefont {S.}~\bibnamefont
  {Weiss}}\ and\ \bibinfo {author} {\bibfnamefont {R.}~\bibnamefont {Egger}},\
  }\href {\doibase 10.1103/PhysRevB.72.245301} {\bibfield  {journal} {\bibinfo
  {journal} {Physical Review B - Condensed Matter and Materials Physics}\
  }\textbf {\bibinfo {volume} {72}},\ \bibinfo {pages} {245301} (\bibinfo
  {year} {2005})},\ \bibinfo {note} {publisher: American Physical
  Society}\BibitemShut {NoStop}%
\bibitem [{\citenamefont {Kyl\"anp\"a\"a}\ and\ \citenamefont
  {R\"as\"anen}(2017)}]{Kylanpaa2017}%
  \BibitemOpen
  \bibfield  {author} {\bibinfo {author} {\bibfnamefont {I.}~\bibnamefont
  {Kyl\"anp\"a\"a}}\ and\ \bibinfo {author} {\bibfnamefont {E.}~\bibnamefont
  {R\"as\"anen}},\ }\href {\doibase 10.1103/PhysRevB.96.205445} {\bibfield
  {journal} {\bibinfo  {journal} {Phys. Rev. B}\ }\textbf {\bibinfo {volume}
  {96}},\ \bibinfo {pages} {205445} (\bibinfo {year} {2017})}\BibitemShut
  {NoStop}%
\bibitem [{\citenamefont {Anderson}\ \emph {et~al.}(2022)\citenamefont
  {Anderson}, \citenamefont {Gyure}, \citenamefont {Quinn}, \citenamefont
  {Pan}, \citenamefont {Ross},\ and\ \citenamefont
  {Kiselev}}]{anderson_high-precision_2022}%
  \BibitemOpen
  \bibfield  {author} {\bibinfo {author} {\bibfnamefont {C.~R.}\ \bibnamefont
  {Anderson}}, \bibinfo {author} {\bibfnamefont {M.~F.}\ \bibnamefont {Gyure}},
  \bibinfo {author} {\bibfnamefont {S.}~\bibnamefont {Quinn}}, \bibinfo
  {author} {\bibfnamefont {A.}~\bibnamefont {Pan}}, \bibinfo {author}
  {\bibfnamefont {R.~S.}\ \bibnamefont {Ross}}, \ and\ \bibinfo {author}
  {\bibfnamefont {A.~A.}\ \bibnamefont {Kiselev}},\ }\href {\doibase
  10.1063/5.0089350} {\bibfield  {journal} {\bibinfo  {journal} {AIP Advances}\
  }\textbf {\bibinfo {volume} {12}},\ \bibinfo {pages} {065123} (\bibinfo
  {year} {2022})}\BibitemShut {NoStop}%
\bibitem [{\citenamefont {Ercan}\ \emph {et~al.}(2021)\citenamefont {Ercan},
  \citenamefont {Coppersmith},\ and\ \citenamefont
  {Friesen}}]{ercan_strong_2021}%
  \BibitemOpen
  \bibfield  {author} {\bibinfo {author} {\bibfnamefont {H.~E.}\ \bibnamefont
  {Ercan}}, \bibinfo {author} {\bibfnamefont {S.~N.}\ \bibnamefont
  {Coppersmith}}, \ and\ \bibinfo {author} {\bibfnamefont {M.}~\bibnamefont
  {Friesen}},\ }\href {\doibase 10.1103/PhysRevB.104.235302} {\bibfield
  {journal} {\bibinfo  {journal} {Physical Review B}\ }\textbf {\bibinfo
  {volume} {104}},\ \bibinfo {pages} {235302} (\bibinfo {year} {2021})},\
  \bibinfo {note} {publisher: American Physical Society}\BibitemShut {NoStop}%
\bibitem [{\citenamefont {Rahman}\ \emph {et~al.}(2012)\citenamefont {Rahman},
  \citenamefont {Nielsen}, \citenamefont {Muller},\ and\ \citenamefont
  {Carroll}}]{Rahman2012}%
  \BibitemOpen
  \bibfield  {author} {\bibinfo {author} {\bibfnamefont {R.}~\bibnamefont
  {Rahman}}, \bibinfo {author} {\bibfnamefont {E.}~\bibnamefont {Nielsen}},
  \bibinfo {author} {\bibfnamefont {R.~P.}\ \bibnamefont {Muller}}, \ and\
  \bibinfo {author} {\bibfnamefont {M.~S.}\ \bibnamefont {Carroll}},\ }\href
  {\doibase 10.1103/PhysRevB.85.125423} {\bibfield  {journal} {\bibinfo
  {journal} {Physical Review B - Condensed Matter and Materials Physics}\
  }\textbf {\bibinfo {volume} {85}},\ \bibinfo {pages} {125423} (\bibinfo
  {year} {2012})},\ \bibinfo {note} {publisher: American Physical
  Society}\BibitemShut {NoStop}%
\bibitem [{\citenamefont {Leon}\ \emph {et~al.}(2020)\citenamefont {Leon},
  \citenamefont {Yang}, \citenamefont {Hwang}, \citenamefont {Lemyre},
  \citenamefont {Tanttu}, \citenamefont {Huang}, \citenamefont {Chan},
  \citenamefont {Tan}, \citenamefont {Hudson}, \citenamefont {Itoh},
  \citenamefont {Morello}, \citenamefont {Laucht}, \citenamefont
  {Pioro-Ladrière}, \citenamefont {Saraiva},\ and\ \citenamefont
  {Dzurak}}]{Leon2020}%
  \BibitemOpen
  \bibfield  {author} {\bibinfo {author} {\bibfnamefont {R.~C.}\ \bibnamefont
  {Leon}}, \bibinfo {author} {\bibfnamefont {C.~H.}\ \bibnamefont {Yang}},
  \bibinfo {author} {\bibfnamefont {J.~C.}\ \bibnamefont {Hwang}}, \bibinfo
  {author} {\bibfnamefont {J.~C.}\ \bibnamefont {Lemyre}}, \bibinfo {author}
  {\bibfnamefont {T.}~\bibnamefont {Tanttu}}, \bibinfo {author} {\bibfnamefont
  {W.}~\bibnamefont {Huang}}, \bibinfo {author} {\bibfnamefont {K.~W.}\
  \bibnamefont {Chan}}, \bibinfo {author} {\bibfnamefont {K.~Y.}\ \bibnamefont
  {Tan}}, \bibinfo {author} {\bibfnamefont {F.~E.}\ \bibnamefont {Hudson}},
  \bibinfo {author} {\bibfnamefont {K.~M.}\ \bibnamefont {Itoh}}, \bibinfo
  {author} {\bibfnamefont {A.}~\bibnamefont {Morello}}, \bibinfo {author}
  {\bibfnamefont {A.}~\bibnamefont {Laucht}}, \bibinfo {author} {\bibfnamefont
  {M.}~\bibnamefont {Pioro-Ladrière}}, \bibinfo {author} {\bibfnamefont
  {A.}~\bibnamefont {Saraiva}}, \ and\ \bibinfo {author} {\bibfnamefont
  {A.~S.}\ \bibnamefont {Dzurak}},\ }\href {\doibase
  10.1038/s41467-019-14053-w} {\bibfield  {journal} {\bibinfo  {journal}
  {Nature Communications}\ }\textbf {\bibinfo {volume} {11}},\ \bibinfo {pages}
  {1} (\bibinfo {year} {2020})},\ \bibinfo {note} {arXiv: 1902.01550 Publisher:
  Nature Research}\BibitemShut {NoStop}%
\bibitem [{\citenamefont {Ercan}\ \emph {et~al.}(2023)\citenamefont {Ercan},
  \citenamefont {Anderson}, \citenamefont {Coppersmith}, \citenamefont
  {Friesen},\ and\ \citenamefont {Gyure}}]{ercan_multielectron_2023}%
  \BibitemOpen
  \bibfield  {author} {\bibinfo {author} {\bibfnamefont {H.~E.}\ \bibnamefont
  {Ercan}}, \bibinfo {author} {\bibfnamefont {C.~R.}\ \bibnamefont {Anderson}},
  \bibinfo {author} {\bibfnamefont {S.~N.}\ \bibnamefont {Coppersmith}},
  \bibinfo {author} {\bibfnamefont {M.}~\bibnamefont {Friesen}}, \ and\
  \bibinfo {author} {\bibfnamefont {M.~F.}\ \bibnamefont {Gyure}},\ }\href
  {\doibase 10.48550/arXiv.2303.02958} {\enquote {\bibinfo {title}
  {Multielectron dots provide faster {Rabi} oscillations when the core
  electrons are strongly confined},}\ } (\bibinfo {year} {2023}),\ \bibinfo
  {note} {arXiv:2303.02958 [cond-mat, physics:quant-ph]}\BibitemShut {NoStop}%
\end{thebibliography}%

% \begin{center}
% \large\textbf{Supporting Information}\par
% \end{center}

\end{document}